\documentclass[]{aastex631}

\received{October 12, 2023}
\revised{December 6, 2023}
\submitjournal{ApJ}

\begin{document}

\title{The Width of Magnetic Ejecta Measured Near 1 au: Lessons from STEREO-A Measurements in 2021--2022}

\author[0000-0002-1890-6156]{No\'{e} Lugaz}
\affiliation{Institute for the Study of Earth, Oceans, and Space, University of New Hampshire, Durham, NH, USA}

\author[0000-0002-5996-0693]{Bin Zhuang}
\affiliation{Institute for the Study of Earth, Oceans, and Space, University of New Hampshire, Durham, NH, USA}

\author[0000-0002-5681-0526]{Camilla Scolini}
\affiliation{Institute for the Study of Earth, Oceans, and Space, University of New Hampshire, Durham, NH, USA}
\affiliation{Solar–Terrestrial Centre of Excellence—SIDC, Royal Observatory of Belgium, Brussels, Belgium}

\author[0000-0002-0973-2027]{Nada Al-Haddad}
\affiliation{Institute for the Study of Earth, Oceans, and Space, University of New Hampshire, Durham, NH, USA}

\author[0000-0002-2917-5993]{Charles~J. Farrugia}
\affiliation{Institute for the Study of Earth, Oceans, and Space, University of New Hampshire, Durham, NH, USA}

\author[0000-0002-9276-9487]{R\'eka~M. Winslow}
\affiliation{Institute for the Study of Earth, Oceans, and Space, University of New Hampshire, Durham, NH, USA}

\author[0000-0002-4017-8415]{Florian Regnault}
\affiliation{Institute for the Study of Earth, Oceans, and Space, University of New Hampshire, Durham, NH, USA}

\author[0000-0001-6868-4152]{Christian M\"ostl}
\affiliation{Austrian Space Weather Office, GeoSphere Austria, Graz, Austria}

\author[0000-0001-9992-8471]{Emma~E. Davies}
\affiliation{Austrian Space Weather Office, GeoSphere Austria, Graz, Austria}

\author[0000-0003-3752-5700]{Antoinette~B. Galvin}
\affiliation{Institute for the Study of Earth, Oceans, and Space, University of New Hampshire, Durham, NH, USA}

%% Note that the \and command from previous versions of AASTeX is now
%% depreciated in this version as it is no longer necessary. AASTeX 
%% automatically takes care of all commas and "and"s between authors names.

%% AASTeX 6.31 has the new \collaboration and \nocollaboration commands to
%% provide the collaboration status of a group of authors. These commands 
%% can be used either before or after the list of corresponding authors. The
%% argument for \collaboration is the collaboration identifier. Authors are
%% encouraged to surround collaboration identifiers with ()s. The 
%% \nocollaboration command takes no argument and exists to indicate that
%% the nearby authors are not part of surrounding collaborations.

%% Mark off the abstract in the ``abstract'' environment. 
\begin{abstract}
Coronal mass ejections (CMEs) are large-scale eruptions with a typical radial size at 1~au of 0.21~au but their angular width in interplanetary space is still mostly unknown, especially for the magnetic ejecta (ME) part of the CME. We take advantage of STEREO-A angular separation of 20$^\circ$--60$^\circ$ from the Sun--Earth line from October 2020 to August 2022, and perform a two-part study to constrain the angular width of MEs in the ecliptic plane: a) we study all CMEs that are observed remotely to propagate between the Sun--STEREO-A and the Sun--Earth lines and determine how many impact one or both spacecraft {\it in situ}, and b) we investigate all {\it in situ} measurements at STEREO-A or at L1 of CMEs during the same time period to quantify how many are measured by the two spacecraft. A key finding is that, out of 21 CMEs propagating within 30$^\circ$ of either spacecraft, only four impacted both spacecraft and none provided clean magnetic cloud-like signatures at both spacecraft. Combining the two approaches, we conclude that the typical angular width of a ME at 1~au is $\sim$ 20$^\circ$--30$^\circ$, or 2--3 times less than often assumed and consistent with a 2:1 elliptical cross-section of an ellipsoidal ME. We discuss the consequences of this finding for future multi-spacecraft mission designs and for the coherence of CMEs.

\end{abstract}

%% Keywords should appear after the \end{abstract} command. 
%% The AAS Journals now uses Unified Astronomy Thesaurus concepts:
%% https://astrothesaurus.org
%% You will be asked to selected these concepts during the submission process
%% but this old "keyword" functionality is maintained in case authors want
%% to include these concepts in their preprints.

%% From the front matter, we move on to the body of the paper.
%% Sections are demarcated by \section and \subsection, respectively.
%% Observe the use of the LaTeX \label
%% command after the \subsection to give a symbolic KEY to the
%% subsection for cross-referencing in a \ref command.
%% You can use LaTeX's \ref and \label commands to keep track of
%% cross-references to sections, equations, tables, and figures.
%% That way, if you change the order of any elements, LaTeX will
%% automatically renumber them.
%%
%% We recommend that authors also use the natbib \citep
%% and \citet commands to identify citations.  The citations are
%% tied to the reference list via symbolic KEYs. The KEY corresponds
%% to the KEY in the \bibitem in the reference list below. 

\section{Introduction} \label{sec:intro}

The properties of coronal mass ejections (CMEs) are known primarily from remote observations, such as coronagraphic and heliospheric imagers, and from {\it in situ} measurements from interplanetary probes that directly measure their magnetic field and plasma properties. By combining hundreds of {\it in situ} measurements, we have learned that the average radial size of a CME at 1~au is $\sim 0.30$~au \citep[]{Richardson:2010}, which can be decomposed into $\sim$ 0.09~au for the dense sheath region that is present in $\sim$80\% of CMEs \citep[]{Salman:2021} and $\sim$ 0.21~au \citep[]{Lepping:2005,Kilpua:2017} for the magnetic ejecta (sometimes referred to as magnetic obstacle). 
In this manuscript, we use the term CME to describe the whole eruption whether it is observed remotely or measured {\it in situ}. The term ICME has been used in the past to refer to the ``interplanetary counterpart'' of a CME%, but some authors use it to refer to the entire eruption while others use this term to only refer to the magnetic ejecta
. Because CMEs are now routinely measured {\it in situ} by Parker Solar Probe \citep[]{Fox:2016,Raouafi:2023} and Solar Orbiter \citep[]{Mueller:2020} at distances where they are also observed remotely by coronagraphs or heliospheric imagers, distinguishing between ICMEs and CMEs is meaningless. The term magnetic ejecta (ME) is meant to refer to the typically low-$\beta$, magnetically dominated ejecta portion of a CME and is favored as it has been used for decades \citep[]{Cane:1998,Lugaz:2012b}. While some authors prefer the term ``magnetic obstacle'', in our opinion, it gives an impression that CMEs are ``obstacles'' whereas in fact they are propagating and expanding structures. Magnetic clouds, which have well-defined properties as described first in \citet{Burlaga:1981} are a special case of MEs, of which they represent 30--40\% of the measurements near 1~au \citep[]{Jian:2018,Richardson:2010}.

The radial properties of CMEs and MEs are relatively well understood; in addition to their size, we know the average radial speed, radial expansion speed \citep[]{Gopalswamy:2015,Jian:2018}. The same is, however, not true for CME properties in the non-radial direction. Some recent work has focused on the magnitude of non-radial flows within CMEs and MEs \citep[]{AlHaddad:2022}, but a key aspect is the ME longitudinal extent. The latitudinal extent of CMEs and MEs can be estimated from remote observations from coronagraphs for the bright front and dark cavity. Estimates for the bright front have been 45--70$^\circ$ \citep[]{Hundhausen:1993,StCyr:2000, Yashiro:2004}, excluding partial and full halo. There has not been any dedicated statistical work on the latitudinal width of the dark cavity, as far as we are aware. Note that, throughout this work, we list the total width rather than the half-width as it is sometimes done for remote observations. 

The angular extent (especially the longitudinal width) of MEs is a critical physical characteristic, both to develop more accurate models and theories of CMEs as well as  for space weather forecasting to know a) how far from the Sun--Earth line measurements can be made that are useful for CME forecasts, and b) how far from the Sun--Earth line a CME must be propagating to impact Earth. Regarding CME models, we note that, under the flux rope paradigm, if the CME leg-to-leg angular extent is comparable to the non-radial cross-section, flux rope models would need to be significantly modified to account for the CME major radius of curvature and the magnetic flux balance within a ``thick'' flux rope. The angular extent is also a critical quantity to investigate ME coherence, an area of recent focus in CME studies \citep[]{Owens:2017,Scolini:2023}.  

To determine the longitudinal width from remote observations, there are a few possible approaches. Under the assumption that CMEs can be represented by a sphere \citep[the so-called ice cream cone model;][]{Xue:2005}, the longitudinal and latitudinal widths are equal. A more common assumption is that the ME part of CMEs can be represented as an axially symmetric highly twisted magnetic flux rope with a ``croissant'' shape \citep[see, for example, the Graduated Cylindrical Shell model of][]{Thernisien:2009}. In-situ studies have revealed that the inclination of MEs (a parameter that affects the latitudinal and longitudinal widths) is more or less random \citep[]{Janvier:2015}, i.e. there is an equipartition of inclination angles. Under the assumption that the number of CMEs measured remotely is high enough, the average longitudinal width should be the same as the average latitudinal width. 

For individual events, or small statistics, fitting and reconstruction techniques can be used to determine the CME angular width from remote observations. \citet{THoward:2012} analyzed the 2008 December 12 CME and found latitudinal widths of 39$^\circ$ and 31$^\circ$ for the bright front and dark cavity, respectively, and longitudinal widths of 56$^\circ$ and 24$^\circ$. For the same event, \citet{Byrne:2010} found an angular width of the bright front (assumed to be spherically symmetric) to increase with distance from 45$^\circ$ to 60$^\circ$ from 0.05 to 0.2~au. \citet{Lugaz:2010b} developed a stereoscopic method which can return the CME longitudinal width, and it was found for two events that the longitudinal width decreases from 90$^\circ$ to 70$^\circ$ as the CME propagates to 0.5~au. \citet{Rodriguez:2011} performed the analysis of 26 CMEs using the Graduated Cylindrical Shell \citep[GCS, see][]{Thernisien:2009,Thernisien:2011} model during the early years of the STEREO missions (late 2007 to mid 2008). The average reconstructed longitudinal width was found to be 51$^\circ$ and the average latitudinal width was 36$^\circ$. \citet{Zhao:2017} studied 33 CMEs with the GCS model and found an average width of 48$^\circ$ (from leg to leg, i.e. ignoring the thickness of the CME). The same average value was found by \citet{Martinic:2022} for 22 CMEs. Lastly, the KINCAT catalog of the Advanced Forecast For Ensuring Communications Through Space (AFFECTS) project has 122 CMEs with an average angular size (leg to leg) of 44$^\circ$. 
We conclude this paragraph by noting two important points. First, in the inner heliosphere, researchers often rely on the self-similar expansion model of \citet{Davies:2012} with a width of 60$^\circ$ or 90$^\circ$, which is higher than the typical width of CMEs as reported from these past studies. Second, there is an overall consensus in the community that the CME angular width remains constant as the CME propagates. This is typically attributed to \citet{StCyr:2000} (see their section 8.5) which was a preliminary study of CME lateral expansion in LASCO/C3 field-of-view. Whether or not this holds true for the ME and beyond 0.15~au (the end of LASCO C3 field-of-view) has not been investigated. 

Multiple single-spacecraft {\it in situ} measurements combined with fitting techniques can be used statistically to derive the typical shape of MEs and CMEs \citep[]{Janvier:2014a,Demoulin:2016}, however, this requires one to assume the CME and ME angular extents. In those works, the authors assume extents of 70$^\circ$ and 60$^\circ$ respectively for the CME and ME. Without such an assumption, the analysis of the distribution of impact parameters resulted in a determination of the CME cross-section to have an aspect ratio of about 3:1. We note that this corresponds to a longitudinal extent of $\sim 35^\circ$ based on the radial size of 0.21~au for a highly inclined flux rope. 
Multi-spacecraft measurements provide the only means by which to determine the longitudinal extent of MEs. \citet{Cane:1998} investigated multi-spacecraft measurements from Helios and noted that ``[except for a potential event measured at 53$^\circ$], there was only one case in which an ejecta was seen at two spacecraft separated by more than 40$^\circ$. In addition, in a number of cases, two spacecraft were separated by less than 40$^\circ$, but an ejecta was seen at only one spacecraft.'' \citet{Kilpua:2011} provided an overview of multi-spacecraft measurements both before the launch of STEREO as well as during the first two years of the missions when the two STEREO spacecraft were separated by 0--45$^\circ$. This includes the well studied November 2007 event \citep[]{Farrugia:2011} with three spacecraft measurements of the same ME over 40.8$^\circ$ separations, but also two cases in December 2007 with only one clear spacecraft measurement of the ME (but potentially inconclusive measurements at a second spacecraft) for total separations of $\sim 43$--44$^\circ$, indicating that two spacecraft separated by 20--25$^\circ$ may not observe the same ME.

One of the most in-depth studies of CME longitudinal extent was performed by \citet{Good:2016} using a database of more than 400 CMEs measured in the inner heliosphere by MESSENGER, Venus Express, {\it Wind}/ACE and STEREO-A and STEREO-B. Their key finding was that there was a probability of $\sim 80$\% for a ME to be measured by two spacecraft separated by less than 15$^\circ$ but it was $\sim 14$\% for separations of 30--60$^\circ$, and no MEs were clearly observed by two spacecraft separated by more than 60$^\circ$. This study has the significant advantage of relying on a very large number of CMEs, but has two main limitations: i) the spacecraft in the innermost heliosphere (MESSENGER and Venus Express) only had magnetic field measurements, and no plasma measurements, which means that the identification relied fully on the magnetic field properties, ii) the spacecraft used in the study were at different heliocentric distances, ranging from 0.31~au (MESSENGER when Mercury is at perihelion) to 1.09~au (STEREO-B at maximum heliocentric distance). This last point means that the CMEs and MEs do evolve for up to a few days between the measurements at the two spacecraft, which could include deflection and non-radial expansion. In addition, the combination of these two limitations means that ensuring that the same ME is measured by two spacecraft separated by more than 15--20$^\circ$ is especially challenging. As such, the number from this study should be taken, in our opinion, as an upper bound. 

%\citet{Rodriguez:2011} performed GCS reconstructions of 26 CMEs measured remotely by the three coronagraphs onboard STEREO-A, STEREO-B and SOHO. Based on the CME direction and width from GCS, they considered that seven CMEs should have been detected at one spacecraft; five of which were indeed measured {\it in situ}. Inversely, one of the 19 CMEs that were not set to impact one of the three spacecraft, impacted one (STEREO-B). They also studied the CME counterparts to the MEs measured {\it in situ}. Overall, they  concluded that GCS can be used adequately to estimate the CME size and direction. 

Since late 2020, STEREO-A has been within 60$^\circ$ from the Sun--Earth line at an heliocentric distance of 0.95--0.98~au and during the ascending phase of solar cycle 25. This provides the opportunity to perform similar studies as those described above but with more CMEs than during 2007--2009 and for two spacecraft at the same heliocentric distance. \citet{Lugaz:2022} described an ME measured by STEREO-A and {\it Wind} in February 2021 while the two spacecraft were separated by 55$^\circ$ in longitude. The CME was propagating almost exactly in-between the Sun--STEREO-A and Sun--Earth lines, giving an angular extent of this event of $\sim 60^\circ$. While the authors conclusively showed that the same ME was measured at the two spacecraft, there were significant global differences in the measurements, such as i) the ME was driving a shock at STEREO-A but not at {\it Wind}, ii) the duration of the ME was about twice shorter at {\it Wind} than at STEREO-A. Many of these differences were due to the interaction of the ME with a high-speed solar wind stream that impacted STEREO-A before the ME but into which the part of the ME that impacted {\it Wind} was engulfed.

In the present work, we take advantage of this return of STEREO-A towards the Sun--Earth line during a time of increased solar activity. Our key goal is to put constraints on the longitudinal extent of MEs through a multi-step approach. We first identify all CMEs that erupted and propagated $\sim 30^\circ$ from the Sun--Earth or Sun--STEREO-A line. We do so from a combination of analyzing remote coronagraphic observations and performing GCS fittings, as described in Section~\ref{data:remote}. We then use existing database and analyze {\it in situ} measurements of CMEs and MEs to determine how many of these CMEs are measured {\it in situ} by none, one or two spacecraft (see Section~\ref{data:insitu}). We also analyze in this section specific examples of CMEs measured {\it in situ} by one or two spacecraft. In Section~\ref{sec:insitu}, we look at all {\it in situ} measurements of CMEs at STEREO-A and {\it Wind} and determine how many of them are measured by the other spacecraft. %Lastly, in section~\ref{sec:L1}, we examine all CMEs that are found to propagate within 30$^\circ$ from the Sun--Earth line to determine how many impact Wind. 
We discuss our results in Section~\ref{sec:discussions}, in particular the consequences for CME coherence and design of future multi-spacecraft missions. We conclude in Section~\ref{sec:conclusions}.

\section{Coronal Mass Ejections Propagating Towards L1 or STEREO-A from November 2020 to August 2022}

\subsection{Remote Observations}\label{data:remote}

We focus on the time period from November 2020 to August 2022 as this corresponds to STEREO-A being 60$^\circ$ to 20$^\circ$ from the Sun--Earth line (always on the east side). During this time period (Nov 2020 -- Aug 2022), the monthly sunspot number was 46.5 $\pm$ 27.5, corresponding to the first half of the rising phase of solar cycle 25. We start our analysis by identifying all partial and full halo CMEs observed by LASCO \citep[]{Brueckner:1995} and STEREO/SECCHI/COR2 \citep[]{Howard:2008}. To do so, we use the SOHO/LASCO catalog \citep[]{Yashiro:2004} and the COR2 CACTUS automatic catalog \citep[]{Robbrecht:2004}. For the LASCO catalog, we start from all CMEs with an angular width larger than 120$^\circ$. For the CACTUS catalog, we start from all events marked as partial or full halo (I, II, III, or IV). This results in 179 CMEs. 

\begin{figure}[t!]
\plotone{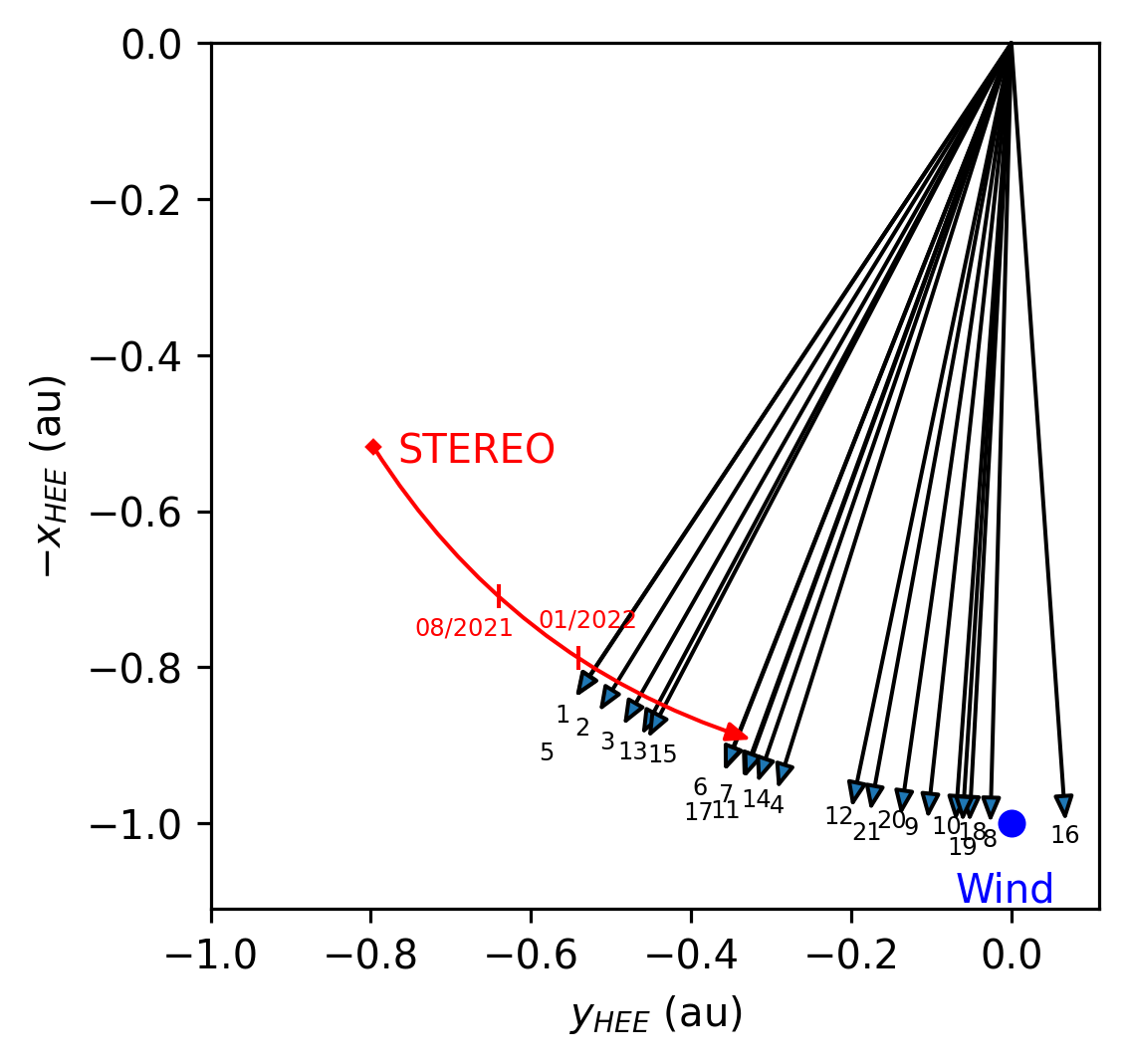}
\caption{Longitudinal direction from the GCS of the 21 CMEs from November 2020 to August 2022 that could impact both STEREO-A and {\it Wind}. The Sun is at the origin (0,0) and this is a top-down view on the ecliptic with each CME average direction shown with an arrow. The plot is in the Heliocentric Earth Ecliptic (HEE) coordinate system. The CMEs are numbered chronologically from November 2020 to August 2022. The position of STEREO-A at the time of CME \#1 is shown with the diamond and the end of the arrow corresponds to CME \#21. In August 2021, STEREO-A was 42$^\circ$ from the Sun--Earth line (CME \#6) and it was 34.5$^\circ$ in January 2022  (CME \#11). Further information about these CMEs can be found in Table~\ref{tab:CatI}.} 
\label{fig:remote}
\end{figure}

We visually inspected LASCO and COR2 movies for each of these events as well as on-disk signatures from SDO/AIA \citep[]{Lemen:2012} to eliminate back-sided events and partial halos from LASCO perspective which are clearly western limb events (i.e. which are propagating far from the Sun--STEREO-A line). This resulted in a list of 51 events that could potentially impact both spacecraft from visual inspection. For these 51 events, two members of the investigation performed independent GCS \citep[using a Python version;][]{gcs_johan} visual fits to the COR2 and C2 images. We primarily focus on the longitudinal direction of the CMEs with respect to the Sun--Earth and Sun--STEREO-A lines and we use the average and minimum (away from the Sun--Earth line) direction from the two visual fits to categorize the CMEs into four groups. Cat I (21 events) are those CMEs that could impact both spacecraft as they are found to propagate within 30$^\circ$ of both Sun--spacecraft lines for at least one reconstruction. Cat II (14 events) are those CMEs propagating within 30$^\circ$ from the Sun--Earth line but more than 30$^\circ$ from the Sun--STEREO-A line. Cat III (four events) are those CMEs propagating within 30$^\circ$ from the Sun--STEREO-A line but more than 30$^\circ$ from the Sun--Earth line. Cat IV (12 events) are those events propagating more than 30$^\circ$ from both Sun--spacecraft lines. The number of CMEs in each category is not representative of their true occurrence rate, as there is a bias towards Earth-directed CMEs by using the more extensive LASCO CME catalog as compared to the automatic CACTUS COR2 catalog. Tables~\ref{tab:CatI} and \ref{tab:appendix} list the information about each Cat I CMEs and all other CMEs, respectively. Some examples of the stereoscopic remote observations for CMEs discussed below are shown in the Appendix in Figure~\ref{fig:appendixOne}. We also estimate the CME speed around 10~$R_\odot$ based on two consecutive GCS reconstructions, where we simply do a first-order differentiation of the height vs.\ time). This is given in the text in the following section, while we use the LASCO/CDAW linear speed in Table~\ref{tab:CatI}.

Figure~\ref{fig:remote} shows a visual sketch of the projected direction of propagation of the 21 Cat I CMEs onto the ecliptic. One of the two GCS fittings for CME \#16 was 7$^\circ$ east of the Sun--Earth line, corresponding to an angular separation of 26$^\circ$ with the Sun--STEREO-A line at that time, even though the average longitudinal direction was 4$^\circ$ west of the Sun--Earth line. The February 2021 CME studied in \citet{Lugaz:2022} corresponds to CME \#2, whereas the 2021 November 2 CME studied in \citet{Regnault:2023b} is CME \#9. 

\subsection{In situ Data and Data Analysis}
We use {\it in situ} measurements from {\it Wind} MFI \citep[]{Lepping:1995} and 3DP \citep[]{Lin:1995} as well as measurements from IMPACT \citep[]{Luhmann:2008} and PLASTIC \citep[]{Galvin:2008} onboard STEREO-A to investigate the impact of the MEs at the two spacecraft and their overall properties. 
In the following sections, we discuss the CME arrival times and properties at the two spacecraft. Determining the fast magnetosonic shock arrival, when one is present, is relatively straight-forward, with the rapid and simultaneous increase in the magnetic field, density, velocity and temperature. Determining the start and end times of the ME is not as straight-forward. The start of the ME is often marked by a decrease in the proton $\beta$, a decrease in temperature below the expected temperature of \citet{Lopez:1987}, an increase in magnetic field strength, a decrease in the magnetic field variability and often a discontinuity in the magnetic field vector. As discussed in \citet{Zurbuchen:2006}, there is no single property which is always measured within MEs. Here, we focus in general on the low $\beta$ region, since it indicates a magnetically dominated period. Choosing a start time that corresponds to a discontinuity in the magnetic field vector often allows to match the start time at the two spacecraft (see for example CME \#5 in the top panels of Figure~\ref{fig:two-SC}). When there is a clear change in the velocity, $\beta$ or magnetic field, we use this as the end time (see for example CME \#12 in the bottom panels of Figure~\ref{fig:two-SC}). Other times, the end time of the ME does not correspond to a clear discontinuity and is harder to determine.

\subsection{Statistics}\label{data:insitu}

To identify the {\it in situ} counterparts to these CMEs, we use the HELIO4CAST catalog \citep[]{Moestl:2020} %, HelioCast} 
for CMEs at STEREO-A and {\it Wind} as well as the catalog of CMEs at ACE by \citet{Richardson:2010}. We visually inspected all {\it in situ} time periods at {\it Wind} and STEREO-A around the expected time of the arrival of the CMEs (typically $\sim$ 2-5 days after the initial appearance in coronagraph images to ensure that we do not miss any potential events). 

\begin{table}[!t]
\caption{Overview of the 21 Cat I CMEs studied here. The columns show, from left to right, the CME ID number, the longitudinal separation between the Sun--Earth and Sun--STEREO-A lines, the date and time of the CME first appearance in LASCO images, the LASCO first-order speed, the GCS longitude separation with L1 ($\phi_{L1}$) and with STEREO-A ($\phi_{STA}$),  the GCS latitude separation with L1 ($\lambda_{L1}$) and with STEREO-A ($\lambda_{STA}$) (a negative sign indicates that the CME propagates south of the Sun-spacecraft line)  and whether there was an impact {\it in situ}.}
    \centering
    \begin{tabular}{|c|c|c|c|c|c|c|c|c|c|}
    \hline
        \textbf{CME \#} & \textbf{$\Delta \phi$ ($^\circ$)} & \textbf{Date} & \textbf{Time (UT)} & \textbf{V (km\,s$^{-1}$)} & \textbf{$\phi_{L1}$($^\circ$)} & \textbf{$\phi_{STA}$($^\circ$)} & \textbf{$\lambda_{L1}$($^\circ$)} & \textbf{$\lambda_{STA}$($^\circ$)} &\textbf{Impact} \\ \hline
        1 & 57 & 1-Dec-20 & 07:12 & 735 & 33 & 24 & 17 & 12 & None  \\ \hline
        2 & 55.6 & 20-Feb-21 & 11:25 & 555 & 31 & 25 & $-10$ & $-15$ & Both \\ \hline
        3 & 54.7 & 20-Mar-21 & 01:25 & 337 & 29 & 26 & 3 & 1 & STA \\ \hline
        4 & 53 & 17-Apr-21 & 17:48 & 353 & 17 & 36 & 10 & 11 & None \\ \hline
        5 & 51 & 22-May-21 & 08:48 & 381 & 33 & 18 & $-6$ & $-1$ & Both \\ \hline
        6 & 42.4 & 23-Aug-21 & 06:48 & 440 & 21 & 21 & 4 & 7 & {\it Wind} \\ \hline
        7 & 42.1 & 26-Aug-21 & 18:48 & 644 & 20 & 23 & 4 & 7 & STA \\ \hline
        8 & 37.4 & 28-Oct-21 & 08:48 & 245 & 2 & 36 & $-34$ & $-36$ & STA \\ \hline
        9 & 37.2 & 2-Nov-21 & 02:48 & 1473 & 6 & 31 & 14 & 11 & {\it Wind} \\ \hline
        10 & 36.2 & 24-Nov-21 & 14:26 & 390 & 4 & 32 & $-17$ & $-21$ & {\it Wind} \\ \hline
        11 & 34.8 & 22-Jan-22 & 10:36 & 278 & 20 & 15 & 26 & 22 & STA \\ \hline
        12 & 34.7 & 29-Jan-22 & 23:36 & 530 & 12 & 23 & 12 & 16 & Both \\ \hline
        13 & 34 & 19-Feb-22 & 17:48 & 274 & 28 & 7 & $-17$ & $-19$ & None \\ \hline
        14 & 33.4 & 16-Mar-22 & 13:25 & 491 & 19 & 15 &13 & 13 & STA \\ \hline
        15 & 33.1 & 25-Mar-22 & 05:48 & 495 & 27 & 6 & $-4$ & $-3$ & None \\ \hline
        16 & 32.9 & 28-Mar-22 & 12:00 & 702 & -4 & 37 & 16 & 16 & {\it Wind} \\ \hline
        17 & 30 & 10-May-22 & 14:36 & 415 & 21 & 9 & $-22$ & $-20$ & None \\ \hline
        18 & 26.2 & 26-Jun-22 & 19:48 & 200 & 3 & 23 & 5 & 8 & STA \\ \hline
        19 & 21.5 & 13-Aug-22 & 18:07 & 606 & 4 & 18 & $-27$ & $-26$ & None \\ \hline
        20 & 21.1 & 15-Aug-22 & 15:24 & 450 & 8 & 13 & $-32$ & $-30$ &None \\ \hline
        21 & 21 & 16-Aug-22 & 02:24 & 495 & 10 & 11 & $-37$ & $-35$ &Both \\ \hline
    \end{tabular}
    \label{tab:CatI}
\end{table}

Overall, from the 21 Cat I CMEs, propagating within 30$^\circ$ of both spacecraft, only four CMEs are  measured at both spacecraft (three events discussed below plus the one from \citet{Lugaz:2022}, namely CMEs \#2, \#5, \#12 and \#21), seven are not measured {\it in situ} at all (CMEs \#1, \#4, \#13, \#15, \#17, \#19 and \#20), six at STEREO-A only (CMEs \#3, \#7, \#8, \#11, \#14 and \#18) and four at {\it Wind} only (CMEs \# 6, \#9, \#10 and \#16). As such 19\% (4 out of 21) of CMEs propagating in-between both spacecraft impacted both, a number that we consider smaller than expected. 
There are other ways to look at these data: (i) 14 of these 21 CMEs (67\%) impacted at least one spacecraft, (ii) there were 18 {\it in situ} measurements of MEs from 42 time periods (21 at STEREO-A and 21 at {\it Wind}) corresponding to potential CME impacts (43\%). 

We consider whether or not CMEs that impact a spacecraft have a smaller angular separation between their direction of propagation (from the GCS model) and the Sun--spacecraft line than those that do not impact a particular spacecraft. We find that there is statistically no difference in the angular separation between the CME direction from GCS and the Sun--spacecraft line for CMEs that impact one spacecraft (18 CME--spacecraft pairs with an average separation of 18.2$^\circ \pm 10.6^\circ$) vs. CMEs that do not impact a spacecraft (24 CME--spacecraft pairs with an average separation 19.4$^\circ \pm 10.9^\circ$). This shows that the study is not just looking at CMEs that are propagating closer to one spacecraft vs. the other.

Hereafter, we discuss specific examples corresponding to different scenarios.  CME \#2 is described in detail in \citet{Lugaz:2022}, where the authors concluded that the ME impacted both STEREO-A and {\it Wind} but with very different {\it in situ} signatures (absence of shock at {\it Wind}, ME duration differs by a factor of two between STEREO-A and {\it Wind}, etc.). Below, we discuss in detail all CME events associated with multi-spacecraft {\it in situ} ME measurements to showcase how different the {\it in situ} measurements look at the two spacecraft. We also discuss one case without any impact, two cases with impact at only one spacecraft, and one complex case.

\subsection{Multi-Spacecraft ME Measurements}

\subsubsection{CME \#5: 2021 May 22 Eruption}

\begin{figure}[t!]
	\centering
	\includegraphics[width=0.4\textwidth]{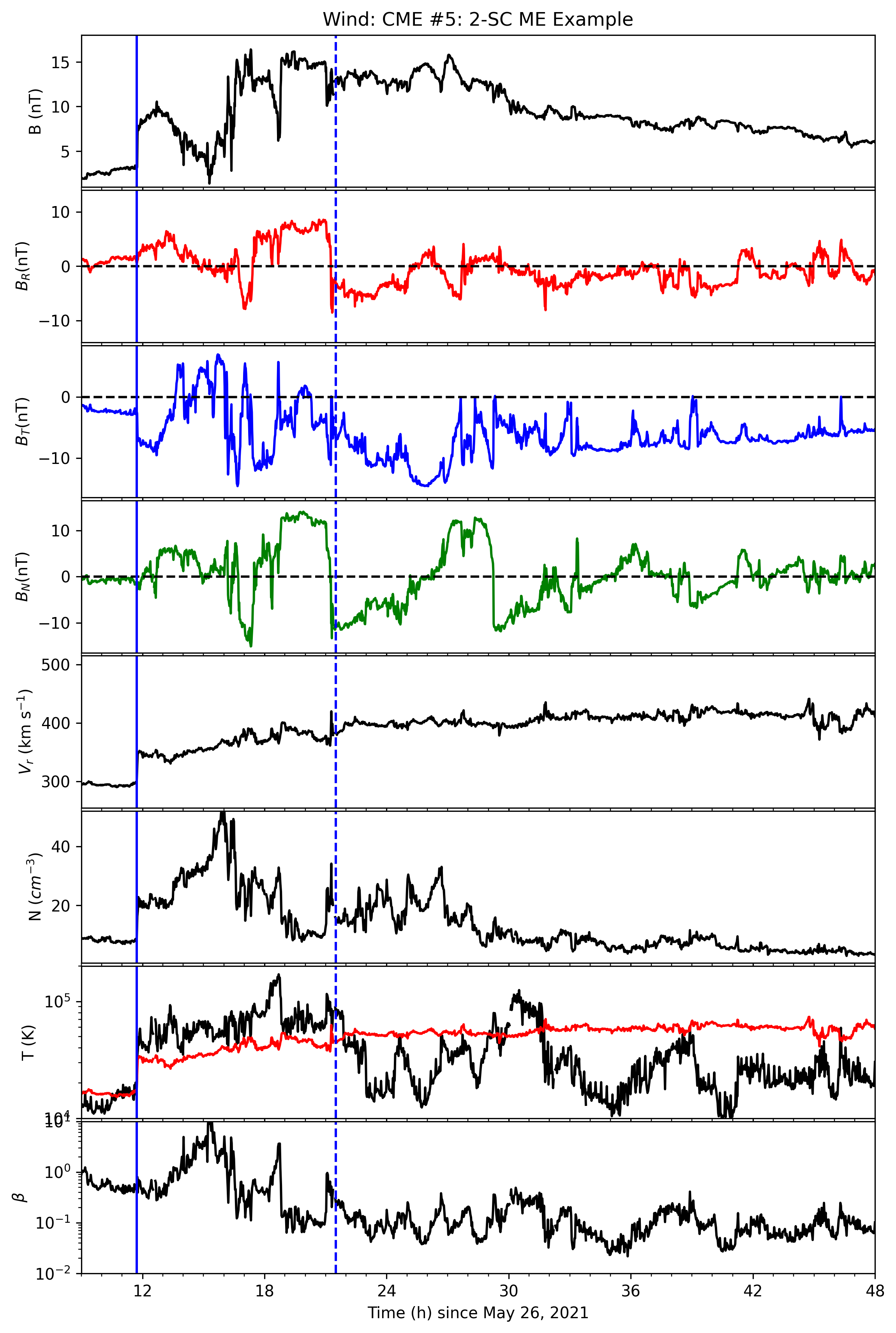}
	\includegraphics[width=0.4\textwidth]{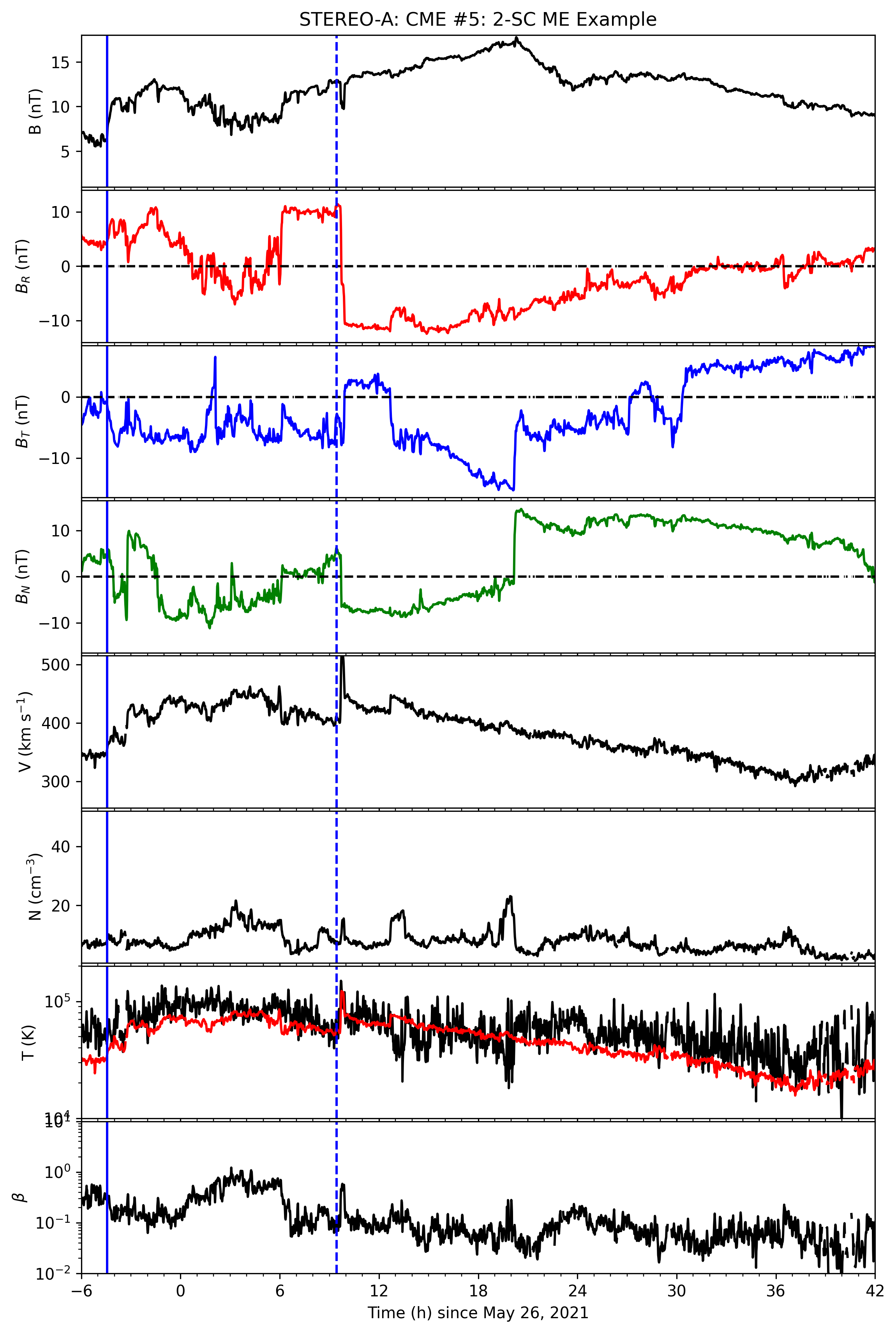}
 	\includegraphics[width=0.4\textwidth]{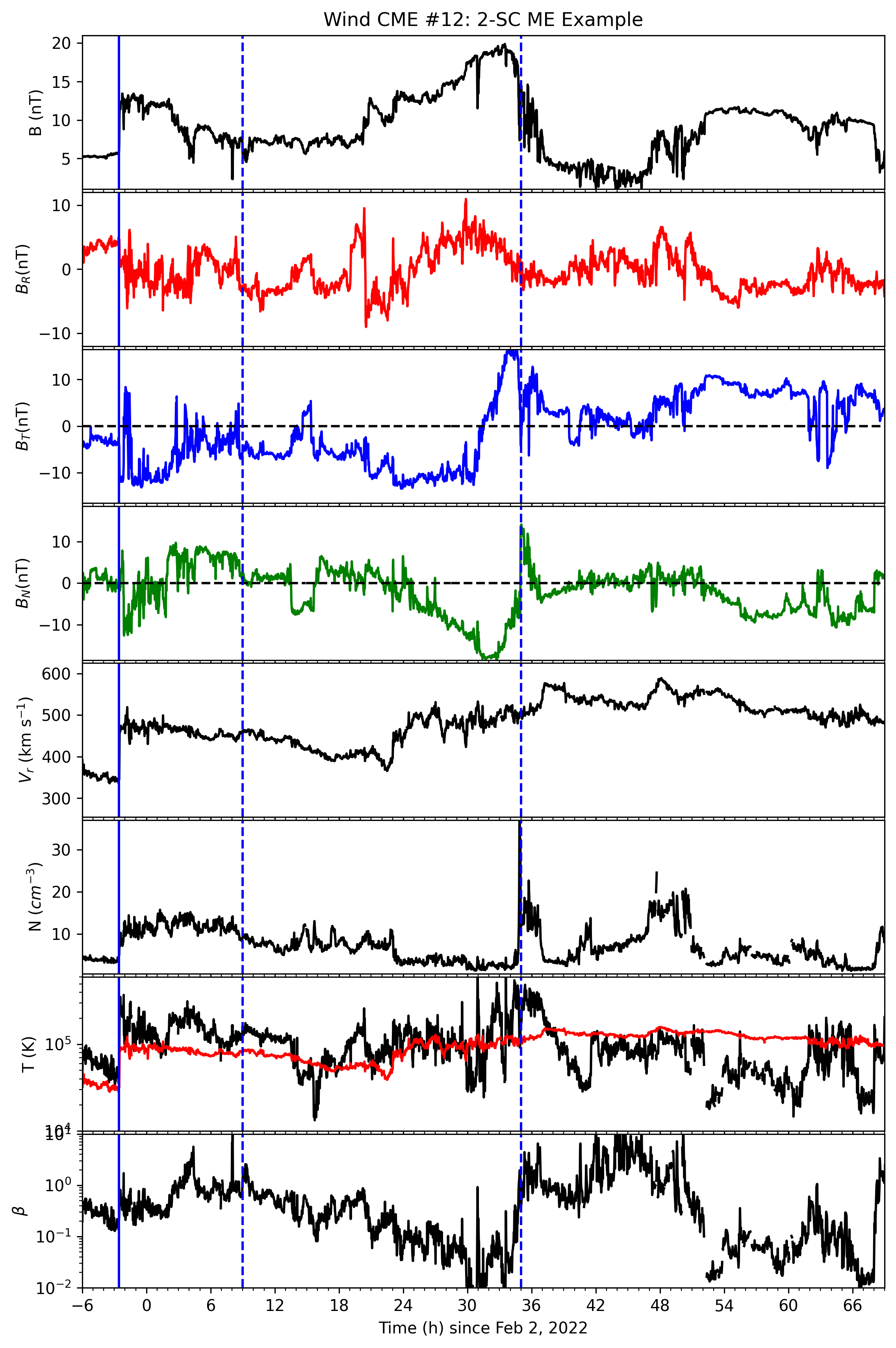}
	\includegraphics[width=0.4\textwidth]{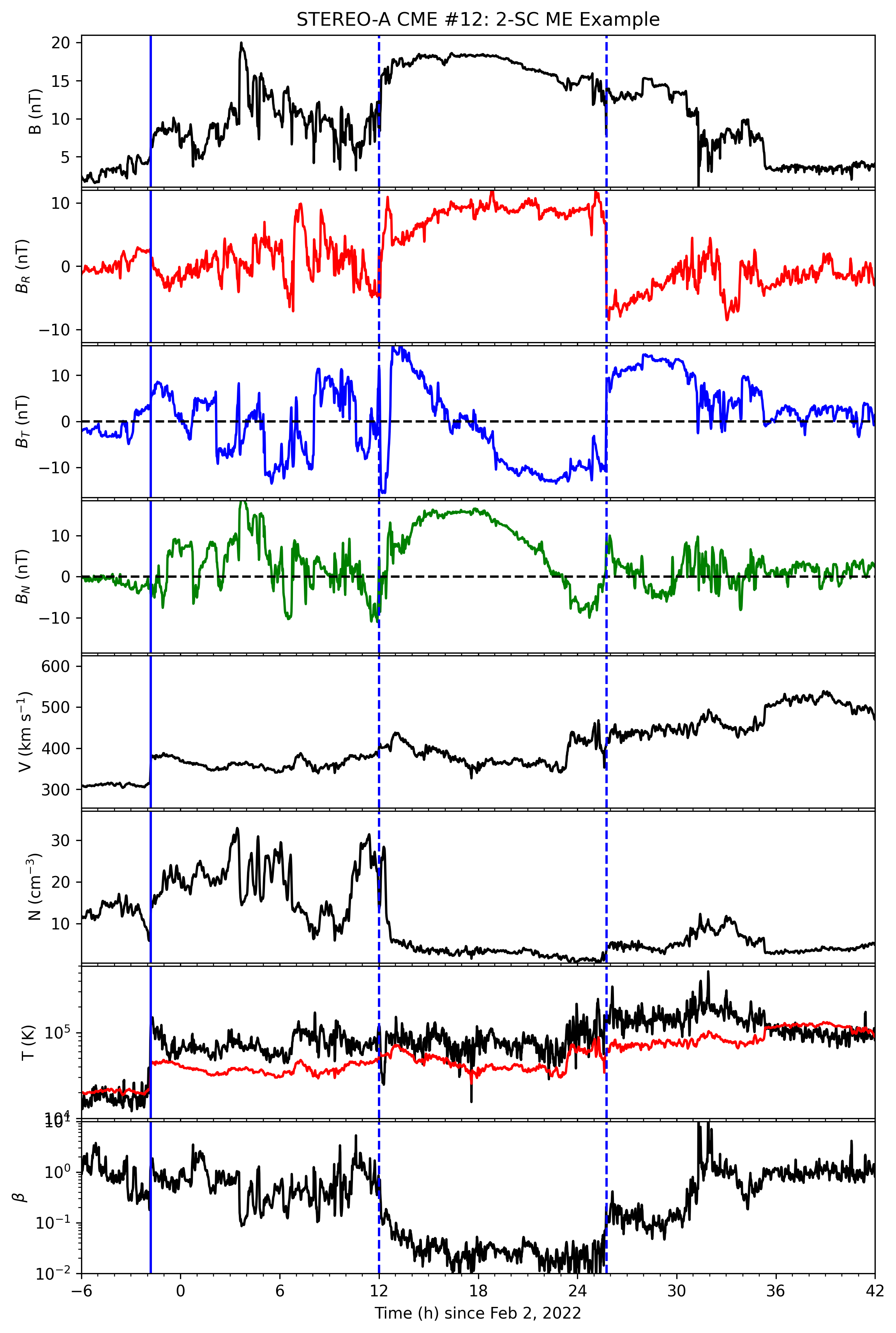}
\caption{Two probable multi-spacecraft measurements of CMEs by {\it Wind} (left) and STEREO-A (right). The top row shows CME \#5 and the bottom shows CME \#12. Each plot shows from top to bottom, the magnetic field strength, the magnetic field components in the RTN coordinate system, the proton velocity, density and temperature (with the expected temperature in red) and proton $\beta$. The plots of {\it Wind} and STEREO-A have the same y-scale for each quantity (but different x-scale). FF shocks are marked with a solid blue vertical line and the ME boundaries with blue dashed vertical lines (the back boundary for CME \#5 is not clear and is not shown).} 
\label{fig:two-SC}
\end{figure}

This CME erupted on 2021 May 22 with a first LASCO image at 08:48 UT, at a time when the L1--Sun--STEREO-A angle was 51$^\circ$. It is associated with an active region which was located on N17E37 at the time of the eruption. The CME is listed as a full halo in the LASCO catalog, but it looks more like a partial eastern halo, which is how it is listed in the CACTUS C2 catalog (with an angular width of 112$^\circ$). GCS reconstruction yields a speed of $\sim$ 450~km\,s$^{-1}$. In STEREO-A/COR2 field-of-view, it appears as a full halo with a dominant western direction. %Due to the number of CMEs occurring in close succession, it is possible that the full halo appearance is due to the contributions of multiple, independent CMEs. 
The two separate GCS reconstructions give a direction of N8E41 and N8E25. We take an average longitudinal direction of 33$^\circ$ $\pm 8^\circ$, and we consider that a direction slightly east of the bisector of the L1--Sun--STEREO-A angle is most likely in light of the relative appearance in LASCO/C2 and SECCHI/COR2 images. (For simplicity, in this manuscript, we list direction east of the Sun-Earth line as positive and west of the Sun-Earth line as negative.) This CME can be expected to impact 1~au at the end of May 25 or early on May 26. Note that STEREO-A heliocentric distance was 0.96~au at that time, which would result in a transit time of $\sim$ 3--4 hours less for a CME speed of 400--450~km\,s$^{-1}$ near 1~au.

A ME can be identified in both STEREO-A and {\it Wind} {\it in situ} measurements starting on May 26 as shown in the top row of Figure~\ref{fig:two-SC}. At {\it Wind}, there is a fast-forward (FF) shock at 11:44 UT followed by a relatively complex ME whose starting time can be chosen as 21:30 UT when the proton temperature goes below the expected proton temperature of \citet{Lopez:1987} at the same time as there is a clear discontinuity in the $R$ and $N$ components of the magnetic field vector. The ME is characterized by enhanced magnetic field strength which peaks in its front half, multiple magnetic field vector rotations and a slowly increasing speed profile. In the database of \citet{Richardson:2010}, it is reported as an MC-0 event, i.e. an event that ``lacks most of the typical features of a magnetic cloud''. There is no specific feature that can define the end boundary, but the magnetic field strength goes back to typical about 30 hours after the beginning of the ME. The end boundary of this CME is not shown in Figure~\ref{fig:two-SC}, as it is unclear.

At STEREO-A, there is a clearer ME preceded by a wave (not a shock) at 19:34 UT on May 25 (shown at time $-4.43$~hours). There is a similar discontinuity in the $R$ and $N$ components of the magnetic field at 9:43 UT on May 26 as the one at {\it Wind}. The ME is characterized by an enhanced magnetic field strength with a ``classical'' peak in the center, a relatively smooth rotation of the magnetic field vector, a decreasing speed profile but a temperature that closely follows the expected temperature. This would probably be categorized as an MC-1 as it lacks the clear low temperature signatures that are most common for true MCs. 

The similarities between the ME measurements at {\it Wind} and STEREO-A are limited to the front half of the ME: the ME start boundary as described above and a period of $\sim$ 8 hours at STEREO-A with $B_N < 0$ and $\sim$ 6 hours at {\it Wind}, followed by a sharp (at STEREO-A) rotation of $B_N$ and $B_T$, which corresponds to the peak of the magnetic field strength at STEREO-A. This time period appears similar at both spacecraft. The period after the sharp rotation in the magnetic field component lasts $\sim 20$ hours at STEREO-A but only 2 hours at {\it Wind}. Instead of a slowly changing magnetic field vector at {\it Wind}, there is a more complex behavior of the magnetic field components. Overall, we conclude from the combination of remote and {\it in situ} measurements that the CME on 2021 May 22 impacted both STEREO-A and {\it Wind}, with a clearer impact at STEREO-A, which is consistent with the visual inspection of the images. The comparison of the magnetic field and plasma measurements at the two spacecraft indicates that, while it is the same ME at both spacecraft, the ME is barely, if at all, coherent at this angular separation. This indicates that the ME angular size is $> 51^\circ$ (the {\it Wind}-STEREO-A separation) but that the coherence length is smaller than this. We also note that, contrary to the event investigated in \citet{Lugaz:2022}, the $B_R$ and $B_T$ components of the magnetic field do not seem to indicate a crossing through a different leg of a flux rope.

\subsubsection{CME \#12: 2022 January 29 Eruption}
This CME erupted on 2022 January 29 with a first LASCO image at 23:36 UT, at a time when the L1--Sun--STEREO-A angle was 34.7$^\circ$. %There is no clear flaring activity associated with the eruption. 
The CME is listed as a full halo in the LASCO catalog with a speed of $\sim$ 500~km\,s$^{-1}$ and it has a clear eastern asymmetry indicating a propagation east of the Sun--Earth line. In STEREO-A/COR2 field-of-view, it appears as a full halo with a dominant western direction. The two separate GCS reconstructions give a direction of N8E10 and N12E13. We take an average longitudinal direction of 11$^\circ$ $\pm 2^\circ$, which is consistent with the slightly more symmetric halo view from L1 than from STEREO-A. This CME can be expected to impact 1~au on or around February 2. 

A FF shock occurs at {\it Wind} at 21:26 UT (time $-2.57$~hours) on February 1 as shown in the bottom left row of Figure~\ref{fig:two-SC}. The next 36 hours are clearly associated with a CME, although the boundaries of the ME are not straight-forward to determine. The database of \citet{Richardson:2010} lists a start time of 16:00 UT on February 2, but the duration of 17.7 hours between the shock and the beginning of the ME is about twice longer than typical for CMEs at 1~au. A start time of the ME around 09:00 UT is also possible based on a discontinuity in $B_T$. While \citet{Richardson:2010} lists this event as a MC-2, we note that the magnetic field strength has a relatively complex profile, the temperature inside the ME is not particularly low, but there is clear rotation of the magnetic field vector starting at 16:00 UT (or 09:00 UT if considering a slightly more complex period) and the speed shows the typical monotonically decreasing profile until 23:00 UT. The ME speed is about 450~km\,s$^{-1}$. The HELIO4CAST database uses a start time of 20:46 UT on February 2 corresponding to a clear discontinuity in $B_R$. Both database list an end time around 10:40 UT on February 3 (34.67~hours). About 13 hours after the end of the ME, there is a second ME with somewhat more typical characteristics. 

At STEREO-A, there is a shock-like feature at 22:11 UT on February 1 (time $-1.82$~hours). There is a clear ME starting between 12:00 and 12:50 UT on February 2 corresponding to a large drop in the proton density and a clear discontinuity in the magnetic field. We consider 01:45 UT on February 3 (25.75~hours) as the end time of the ME, corresponding to an increase in the proton $\beta$ and a clear discontinuity in the magnetic field. %The following 4.5 hours are associated with a separate rotation of the magnetic field vector, and the HELIO4CAST database lists this as a separate CME, but the very short duration makes it unlikely to have an  origin independent from that of the previous ME. 
The ME has a speed of about 360~km\,s$^{-1}$. 

Based solely on the {\it in situ} measurements, there are few, if any, signs that STEREO-A and {\it Wind} measured the same event. However, the beginning of the sheath at the two spacecraft is within 1.5 hours from each other and there is just one CME that propagated in-between the Sun--STEREO-A and Sun--L1 lines at a time and with a speed that match the start time of the event. We consider that it is likely the same event, but this is primarily based on the remote observations and the fact that STEREO-A and {\it Wind} were separated by less than 35$^\circ$ at that time. This would indicate that the ME angular size is $35^\circ$ but the coherence length is significantly less than 35$^\circ$.

\subsubsection{CME \#21: 2022 August 15 Eruption}
There were at least five front-sided full or partial halo CMEs from 2022 August 13 to 17 with the ones on August 13 and 15 potentially propagating in-between the Sun--Earth and Sun--STEREO-A lines (CMEs \#20 and 21 in our list). At the time, STEREO-A was separated by 21--21.5$^\circ$ from the Sun--Earth line. We focus here on the August 15 CME, which has a first appearance in LASCO images at 15:24 UT with an apparent angular width of 125$^\circ$. It originates from an active region which was around S20W01 at that time. As STEREO-A and SOHO were only separated by 21$^\circ$, the GCS reconstruction has more uncertainty, but it appears visually that the CME propagates in-between the two spacecraft, being slightly eastwards from SOHO/LASCO point-of-view and slightly westward from STEREO-A/COR point-of-view. The two GCS reconstructions give direction of S25E07 and S25E09, so that we take an average longitudinal direction of 8$^\circ$ $\pm 1^\circ$. Based on the speed measured in coronagraphs, we can expect an arrival at 1~au on or around August 19.

\begin{figure}[t!]
	\centering
	\includegraphics[width=0.4\textwidth]{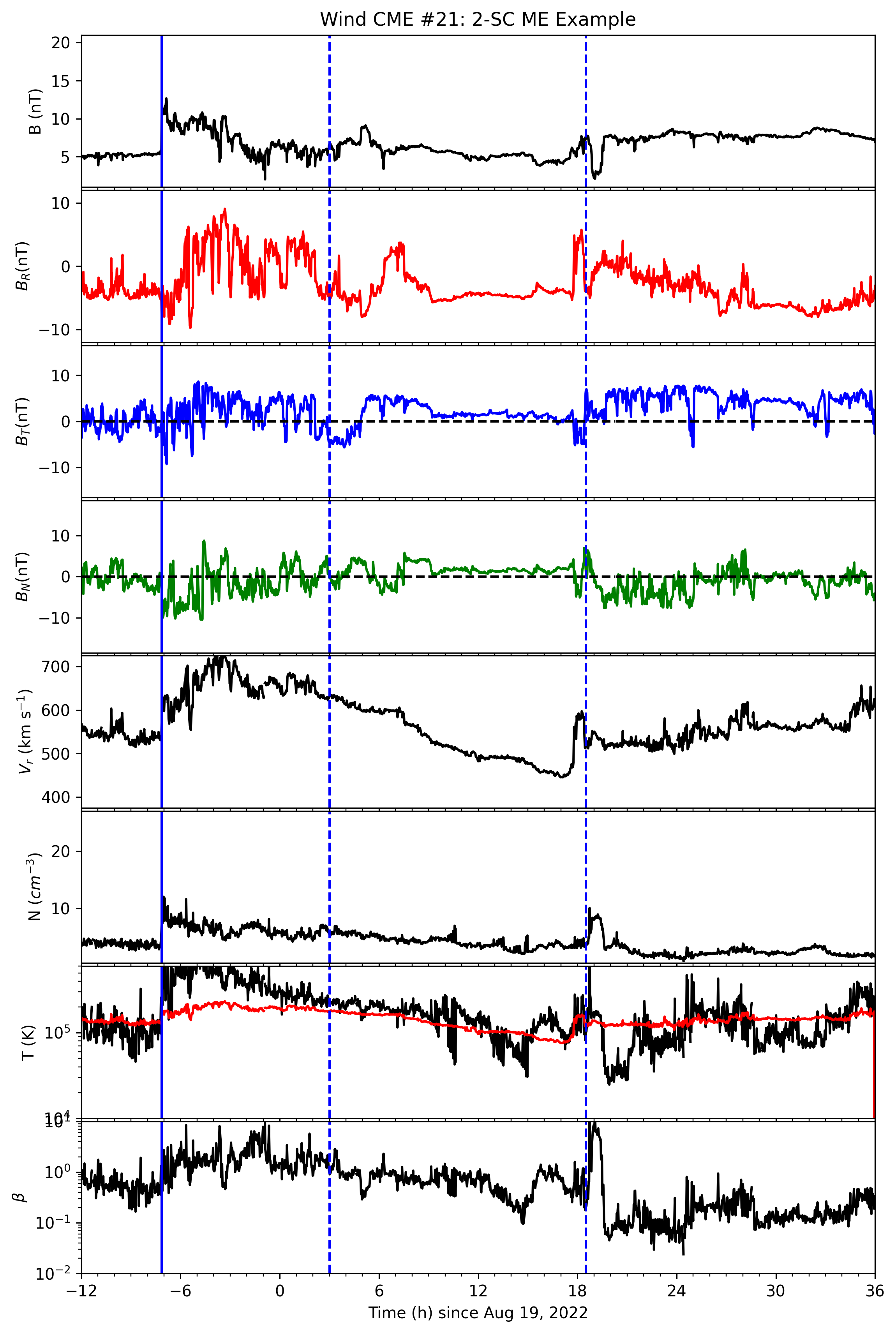}
	\includegraphics[width=0.4\textwidth]{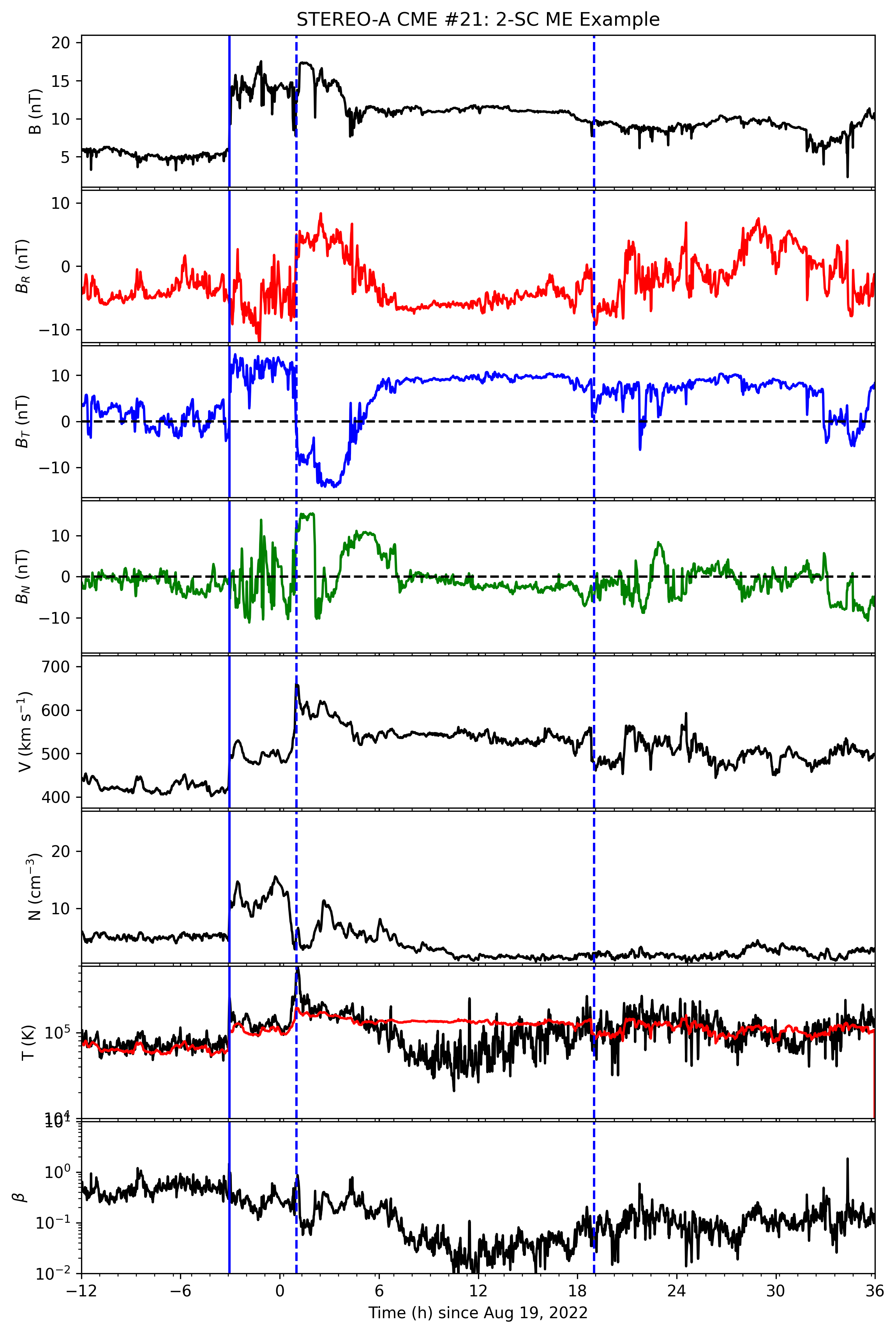}
 	\includegraphics[width=0.4\textwidth]{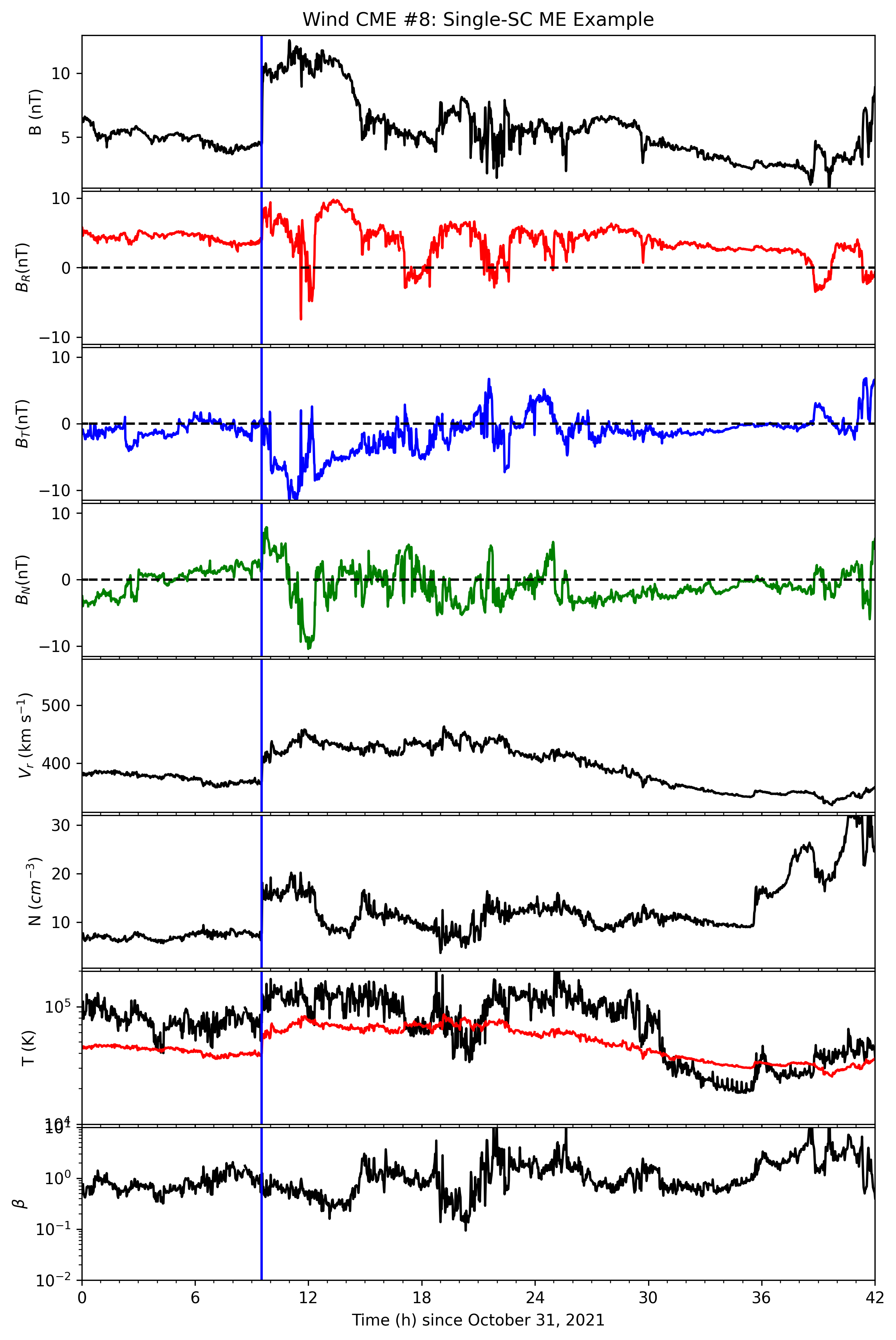}
	\includegraphics[width=0.4\textwidth]{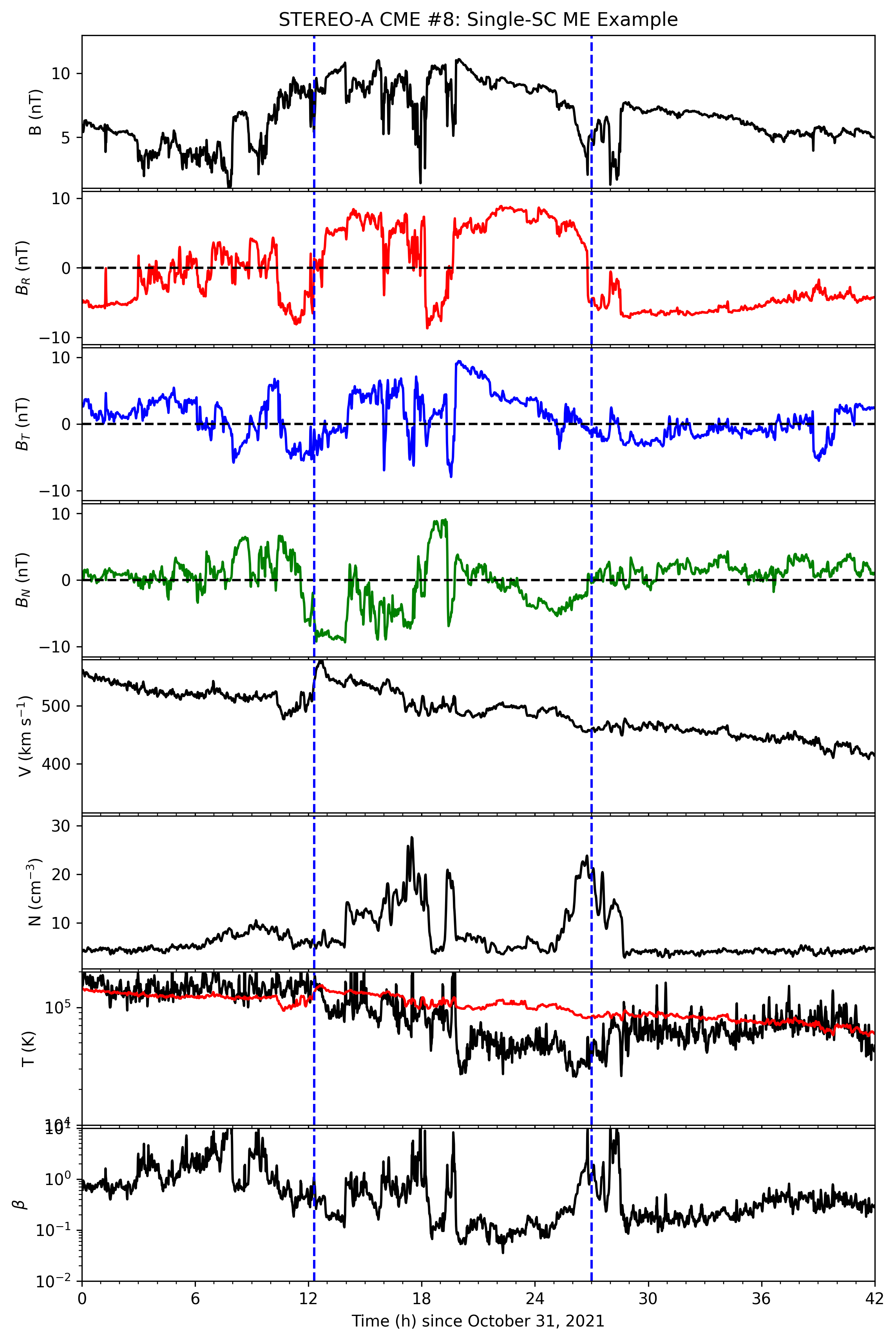}
\caption{{\it Wind} (left) and STEREO-A (right) measurements of a two-spacecraft ME measurements (CME \#21, top) and a complex single-spacecraft ME measurement (CME \#8, bottom, ME at STEREO-A, shock at {\it Wind}). Same format as Figure~\ref{fig:two-SC} but both x-scale and y-scale are the same between STEREO-A and {\it Wind.}} 
\label{fig:one-SC}
\end{figure}

{\it Wind} measures a FF shock at 16:51 UT on August 19 (see top plot of Figure~\ref{fig:one-SC}) followed by a ME that starts around 03:00 UT on August 20, although the list of \citet{Richardson:2010} has a later start time. The end time is at around 18:30 UT on August 20. The ME has relatively weak magnetic field strength ($\sim$~6~nT), no clear rotation of the magnetic field vector, temperature similar to the expected temperature, and not very low proton $\beta$. However, the magnetic field variability is small and there is a clear decreasing speed profile. The ME has an average speed of 550~km\,s$^{-1}$ with an expansion speed of $\sim$ 95~km\,s$^{-1}$. There is no ME listed in the HELIO4CAST catalog but a MC-1 listed in the list of \citet{Richardson:2010}.

STEREO-A measures a FF shock at 20:57 UT on August 19 followed by a ME which starts around 01:00 UT on August 20 (HELIO4CAST lists a 05:16 UT start time) and lasts until 19:00 UT on August 20. The magnetic field strengths in the sheath and in the ME are clearly stronger at STEREO-A than at {\it Wind} (ME field strength of $\sim$ 12~nT). There is a clearer portion of the ME with low proton $\beta$ and a lower than expected proton temperature. %While there is some rotation of the magnetic field vector, it is dominated by a period of about 4--5 hours at the beginning of the ME, which may be part of the sheath (higher density and temperatures). 
The ME has a speed of $\sim$ 550~km\,s$^{-1}$ with almost no expansion. Similar to the previous event, if it were not for the near-simultaneous measurements, the close proximity of L1 and STEREO-A ($\sim 21^\circ$) and the fact that there is a CME propagating in-between the two spacecraft, it is possible that these two events would not be linked. While this might be related to the origin of the CME south of the ecliptic, we note that the STEREO-A measurements differ significantly from those from {\it Wind}. %In addition, the 2003 October 28 CME (the ``Halloween'' event) originated from an active region at S17E03, not fundamentally different from the origin of this eruption. 

\subsection{Single-Spacecraft {\it In Situ} ME Measurements and No Impact}
Every other 17 CME that is found remotely to propagate in-between the Sun--Earth and Sun--STEREO-A lines impacts at most one spacecraft, either {\it Wind} or STEREO-A. We give four examples, one for a detection at STEREO-A only, and one without impact at any spacecraft near 1~au, as well as two events which are harder to interpret.

\subsubsection{CME \#8 2021 October 28 Eruption: A Complex Event}
This CME erupted on 2021 October 28 with a first LASCO image at 8:48 UT, at a time when the L1--Sun--STEREO-A angle was 37.4$^\circ$. GCS reconstructions give a direction of S30E2 for the CME in the corona, so it is directed almost exactly along the Sun--Earth line (but south of the ecliptic) and $\sim 35^\circ$ from the Sun--STEREO-A line. 

There is a potential ME event measured at STEREO-A lasting 14 hours and starting around 12:20 UT on October 31 (see bottom right plot of Figure~\ref{fig:one-SC}). We consider this as a sheath-less  ME with complex magnetic field vector rotations with $B_N$ mostly negative and a magnetic field magnitude around 10~nT. There is a period of colder-than-expected temperatures and low proton $\beta$ for $\sim$ 8 hours in the second half of the event. At a similar time at 9:32 UT on October 31, {\it Wind} measures a clear FF shock followed by weak magnetic field and period of typical proton $\beta$. About 20 hours after the shock, there is a short period ($\sim$ 4 hours) of lower than expected temperature. We consider that this event is a shock-less and sheath-less ME measured at STEREO-A and a ``driverless'' shock measured at {\it Wind}. Whether or not it is associated with the October 28 CME is beyond the scope of this work as well as confirmation that the shock measured at STEREO-A is the same shock as measured at {\it Wind}.

\begin{figure}[t!]
	\centering
	\includegraphics[width=0.4\textwidth]{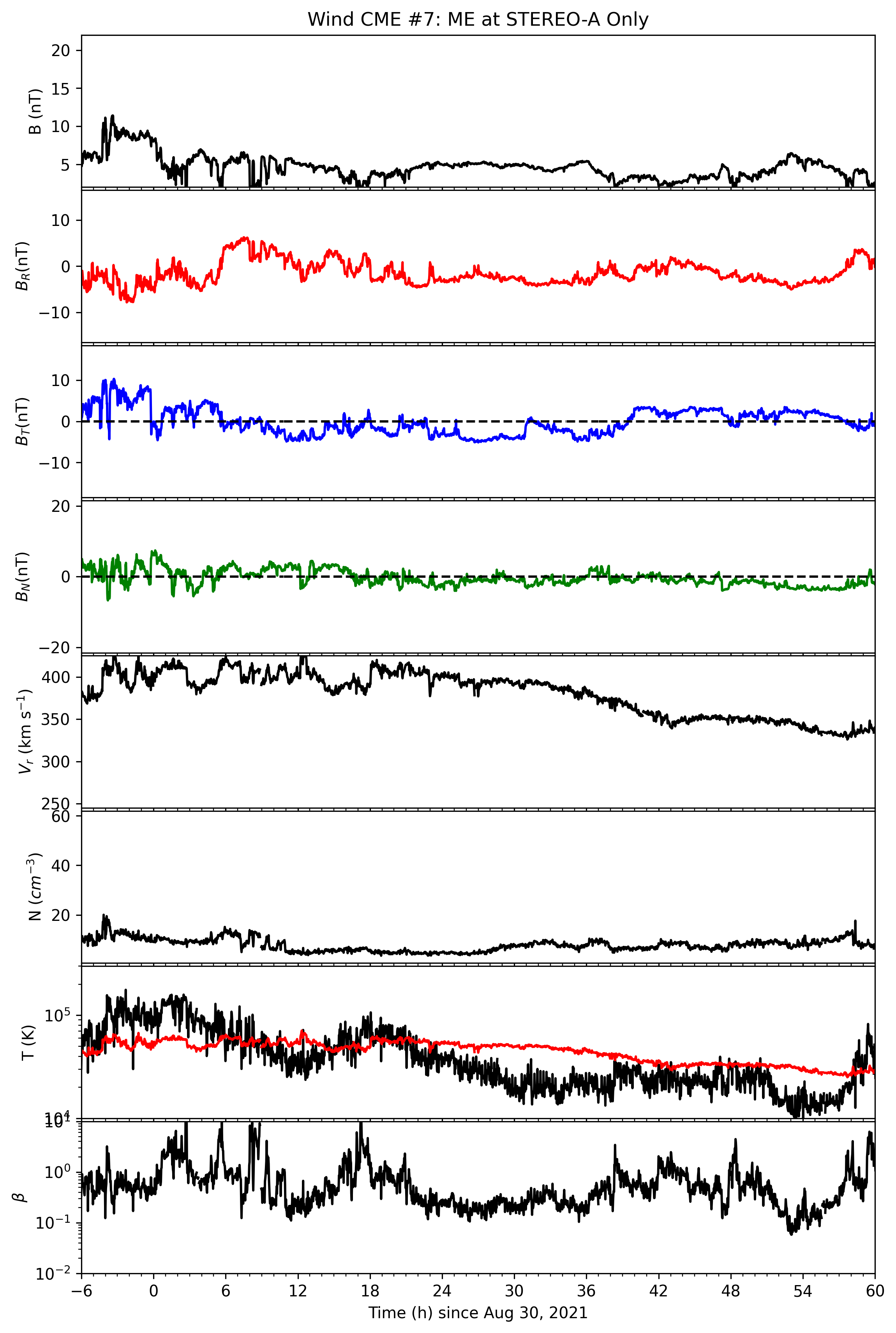}
	\includegraphics[width=0.4\textwidth]{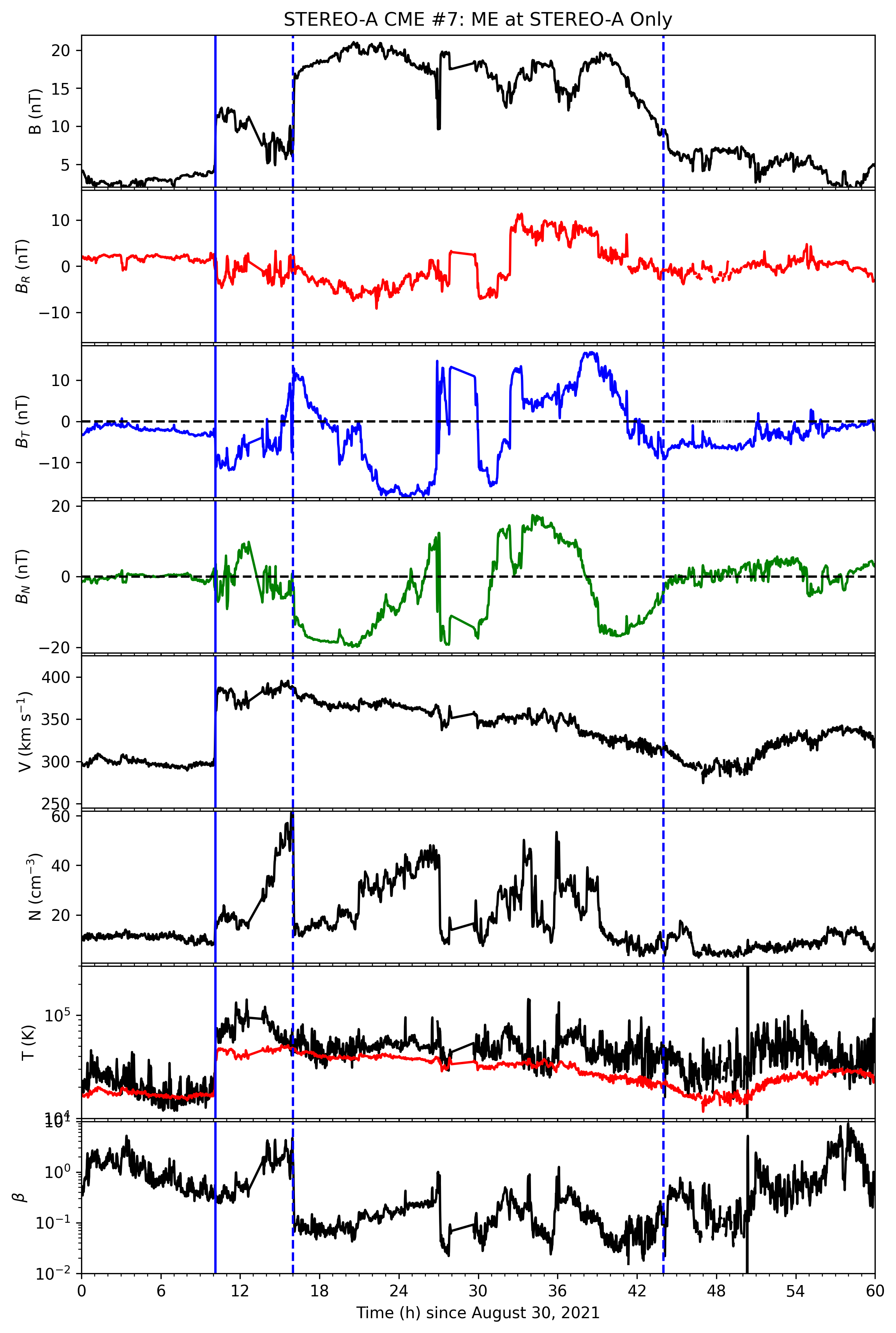}
 	\includegraphics[width=0.4\textwidth]{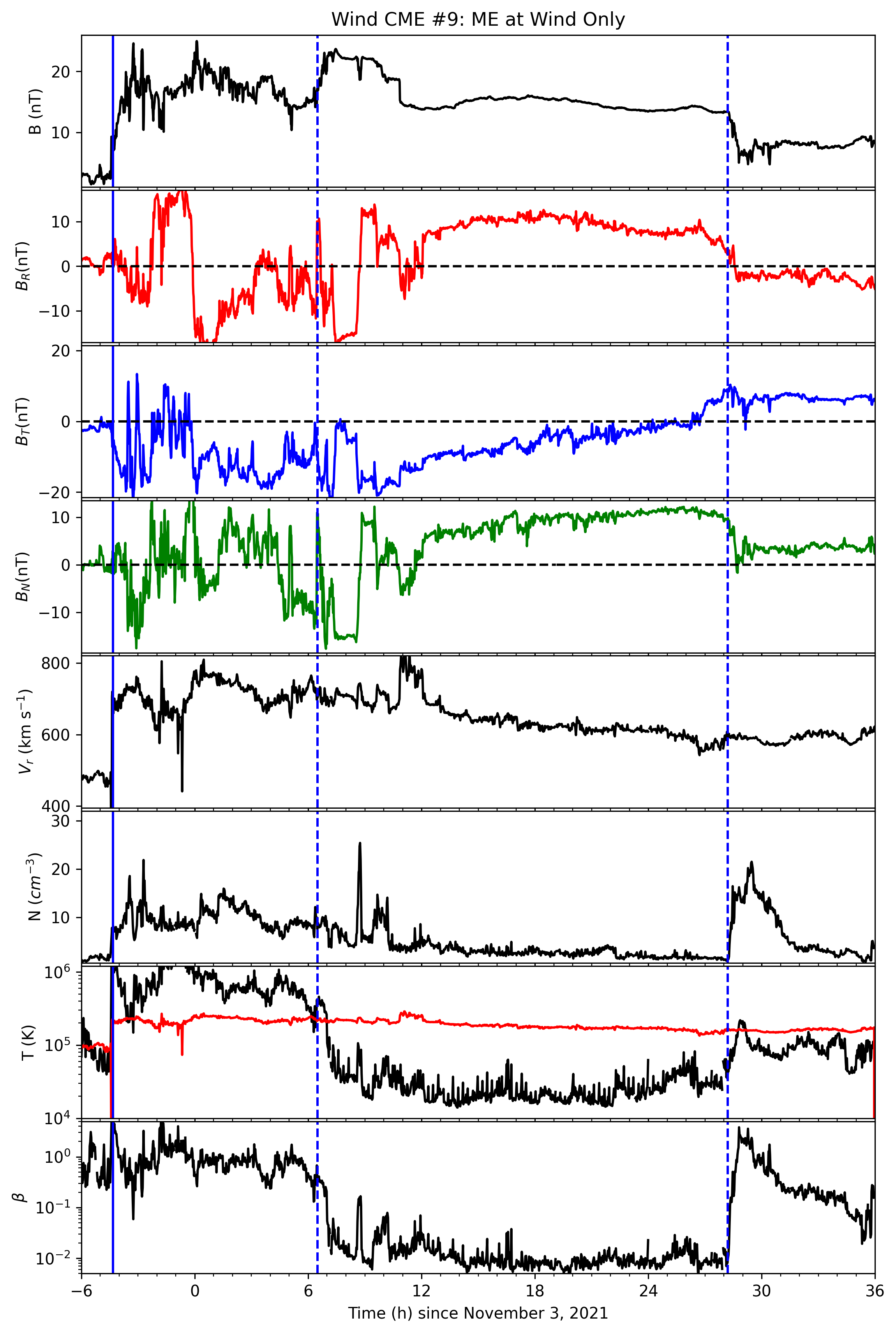}
	\includegraphics[width=0.4\textwidth]{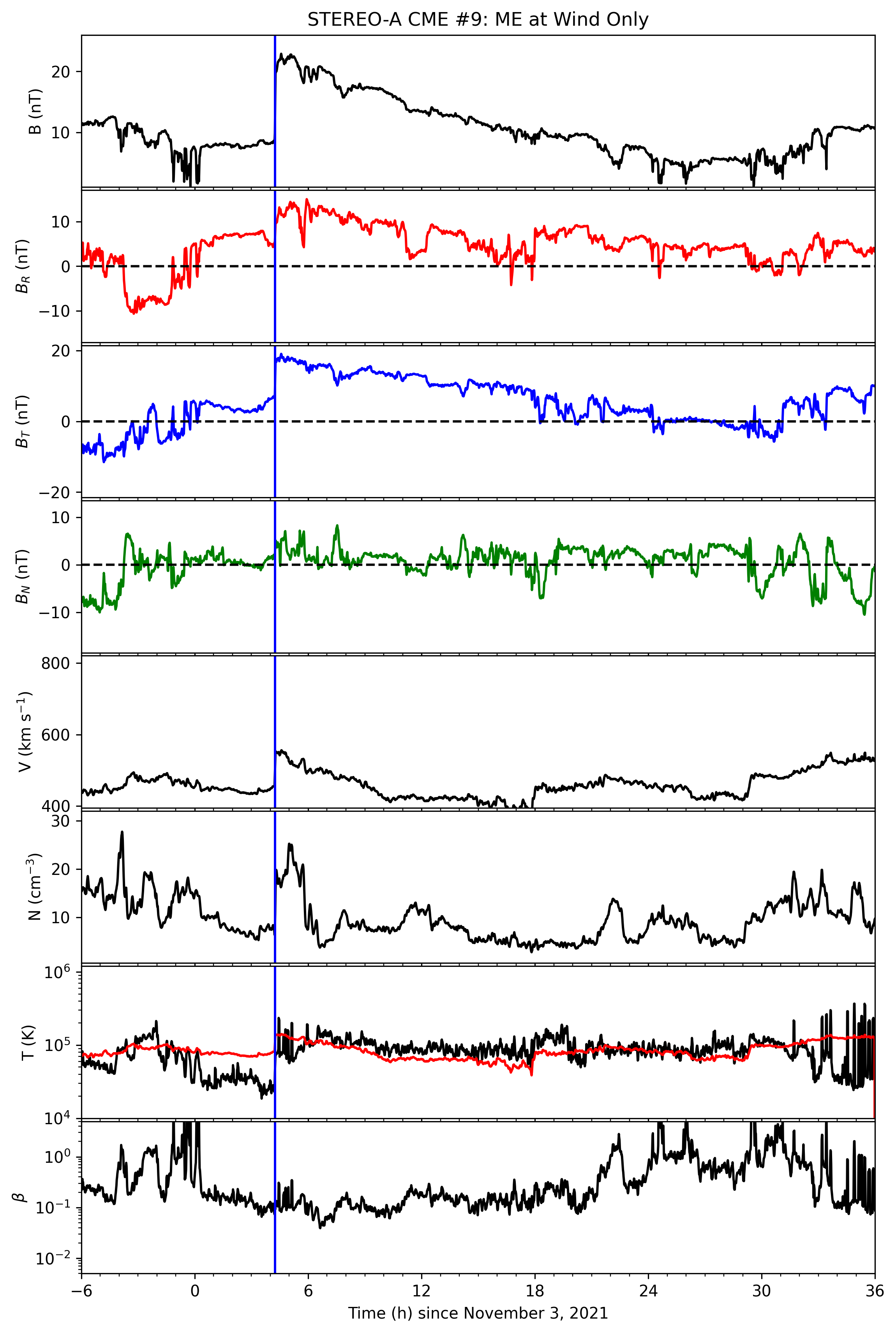}
\caption{{\it Wind} (left) and STEREO-A (right) measurements of a clear single-spacecraft ME measurements: CME \#7 at STEREO-A (top) and CME \#9 at {\it Wind} (bottom).  Same format as Figure~\ref{fig:two-SC}.} 
\label{fig:onetrue-SC}
\end{figure}

\subsubsection{CME \# 7: 2021 August 26 Eruption: Single-spacecraft ME at STEREO-A}
This CME erupted on 2021 August 26 with a first LASCO image at 18:48 UT, at a time when the L1--Sun--STEREO-A angle was 42$^\circ$ and associated with an active region around N20E14. The CME is listed as a full halo in the LASCO catalog with  %but might be better described as a partial halo with a clear eastern asymmetry indicating a propagation east of the Sun--Earth line. It has 
a speed of $\sim$ 750~km\,s$^{-1}$. In STEREO-A/COR2 field-of-view, it appears as a partial halo with a dominant western direction. GCS reconstructions gave a direction of N11E20.

A FF shock occurs at STEREO-A at 10:08 UT on August 30 (see top right plot in Figure~\ref{fig:onetrue-SC}). A ME starts around 16~UT on August 30 with a rotation of the magnetic field vector and a decrease in proton density. The magnetic field and density profiles are complex but the speed is monotonically decreasing for more than 36 hours with an average speed of $\sim$ 350~km\,s$^{-1}$ and an expansion speed $\sim$ 50~km\,s$^{-1}$. The magnetic field strength has an average $\sim 15$~nT with multiple complete rotations of the magnetic field vector. HELIO4CAST lists two MEs separated by about 10 hours but the period in-between has a similar speed profile and we consider that the {\it in situ} measurements are consistent with a multi-MC event as described in \citet{Lugaz:2017}. At the same time at {\it Wind}, there are no clear signatures of any ME. There is a period of the proton temperature being slightly lower than the expected temperature, which corresponds to a period with slightly less variability in the magnetic field vector. However, the magnitude of the magnetic field is $\sim$ 5~nT and the $B_N$ component which has three clear periods of strong negative $B_N$ at STEREO-A is close to 0 throughout this time period at {\it Wind}.

The direction of the CME is clearly in-between STEREO-A and {\it Wind}, the clear signature of a complex event at STEREO-A and the lack of transient signatures at {\it Wind} is confusing for a spacecraft separation of $42^\circ$ and a propagation direction close to the angle bisector. It is possible that the interaction of multiple CMEs may have played a role in the deflection of the CMEs towards the east (STEREO-A) although confirming this is beyond the scope of this study. We note (see Figure~\ref{fig:appendixTwo}) that the CME has a clear component in the ecliptic and a propagation clearly out of the ecliptic cannot be invoked to explain the {\it in situ} measurements. The measurements would indicate that the ME angular size is less than $42^\circ$ but, if no deflection is invoked, the ME angular size would be less than $25^\circ$. 

\subsubsection{CME \#9 2021 November 2 Eruption: Single-spacecraft ME at Wind}
This CME erupted on 2021 November 2 with a first LASCO image at 02:48 UT, at a time when the L1--Sun--STEREO-A angle was 37.2$^\circ$. GCS reconstructions give a direction of S19E6, so directed very close to the Sun--Earth line (but south of the ecliptic) and $\sim 30^\circ$ from the Sun--STEREO-A line. This is a relatively fast CME with a speed of more than 1400~km\,s$^{-1}$.

There is a clear ME at {\it Wind} with a FF shock at 19:40 UT on November 2 (see bottom left plot in Figure~\ref{fig:onetrue-SC}) about 41 hours after the first image of the CME on LASCO. {\it In situ} speeds above 750 ~km\,s$^{-1}$ are measured, both in the sheath and in the front half of the ME which starts around 6:30 UT on November 3. There is a clear rotation of $B_T$ from negative to positive with a positive $B_N$. The ME is characterized by a low temperature, low proton $\beta$ and a relatively clear expansion. STEREO-A measures near the same time a complex event that starts with a discontinuity at 4:15 UT on November 3, about 8.5 hours after the shock at {\it Wind}. There are more complex magnetic field signatures as well as high density and a speed of 450--500~km\,s$^{-1}$. $B_T$ is primarily positive. %Starting on November 4, this is followed  by what looks like a ``driverless'' shock and more disturbed conditions. 
%Because of the long delay between STEREO-A and {\it Wind} measurements, the very different magnetic field profiles and different velocity, we consider that the measurements at STEREO-A are not associated with CME \#9. As such, 
Confirming whether or not the shock is driven by the same CME at {\it Wind} and STEREO-A is beyond the scope of this study. In any case, there is only a ME at {\it Wind}.

%%Note for us: Are we using the term "complex ME" in the same sense as Burlaga use of complex ejecta. I guess not. We mean more like "complicated". In that case we should make an early qualifier on this,I think.
  
\subsubsection{CME \#15: 2022 March 25 Eruption: No Impact}
This CME erupted on 2022 March 25 with a first LASCO image at 05:48 UT, at a time when the L1--Sun--STEREO-A angle was 33$^\circ$ and associated with an active region around S20E36. The CME is listed as a full halo in the LASCO catalog and it has a clear eastern asymmetry indicating a propagation east of the Sun--Earth line. It has a speed of $\sim$ 600~km\,s$^{-1}$. In STEREO-A/COR2 field-of-view, it appears as a full halo with a very slight asymmetry to the west. GCS reconstructions gave a direction of S10E27 with uncertainties of $\pm 4^\circ$. The CME is directed almost exactly towards STEREO-A but because the two spacecraft separation was small, an impact at both spacecraft can be expected. 

\begin{figure}[t!]
	\centering
	\includegraphics[width=0.4\textwidth]{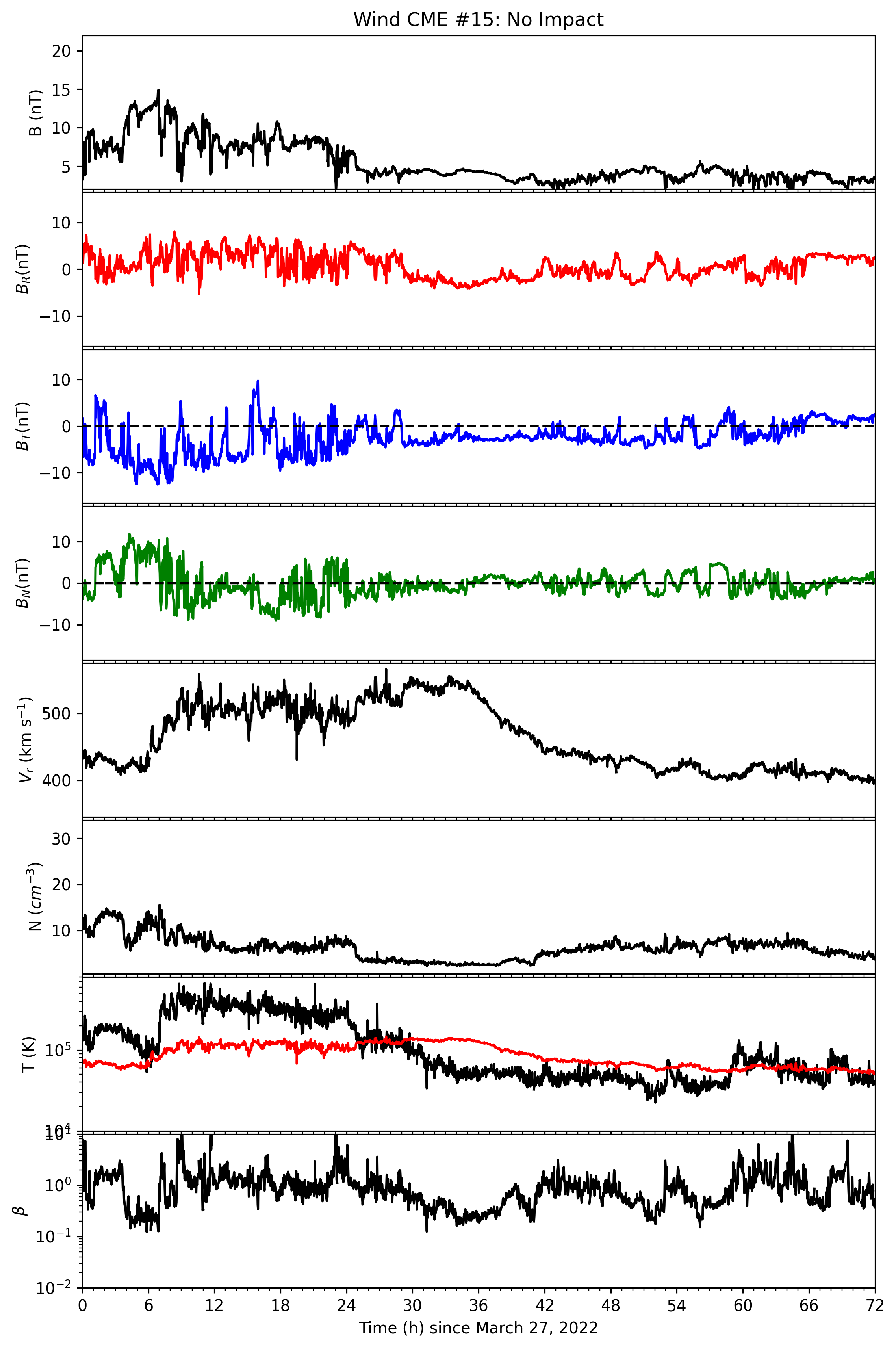}
	\includegraphics[width=0.4\textwidth]{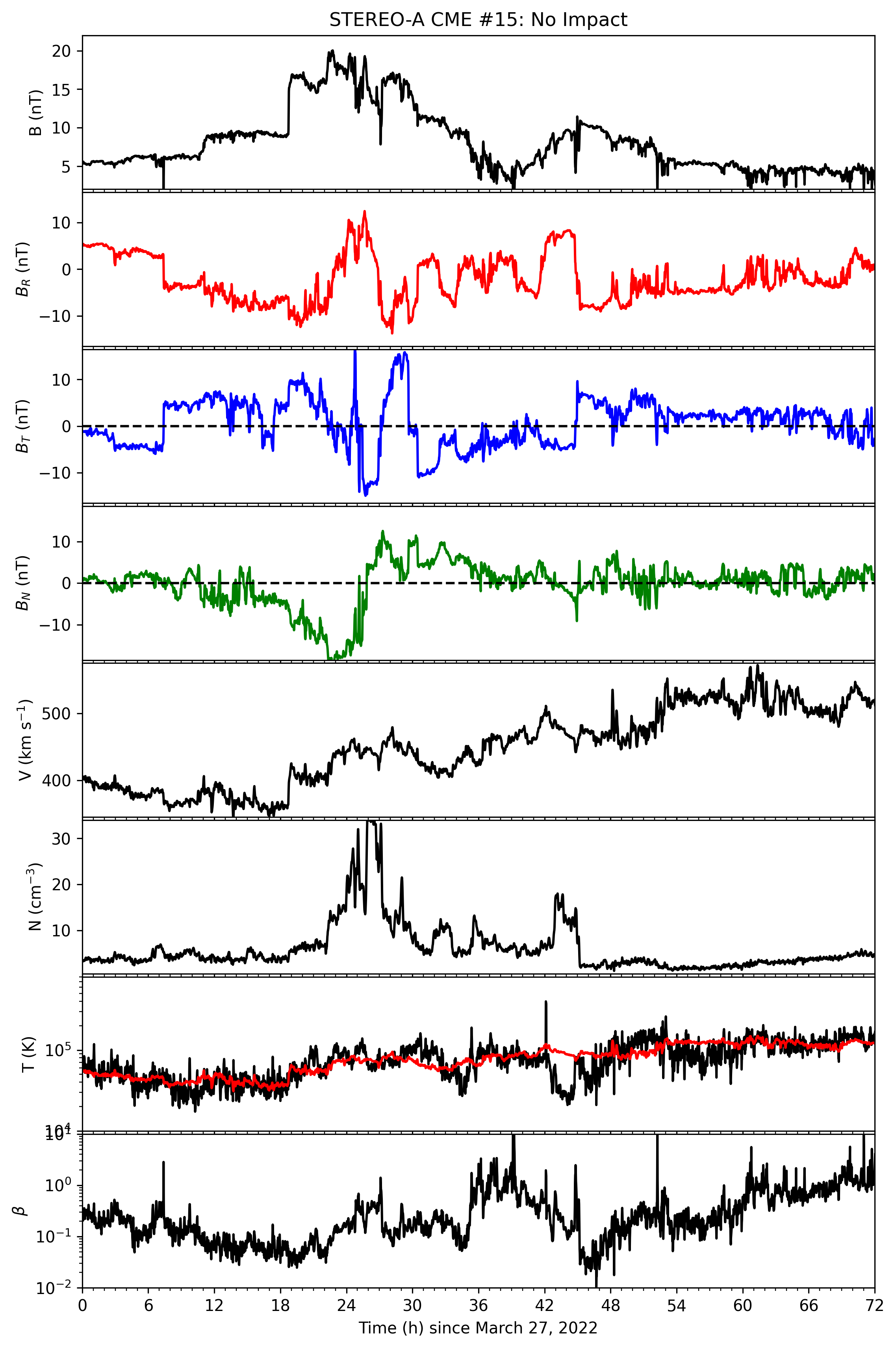}
 	\includegraphics[width=0.4\textwidth]{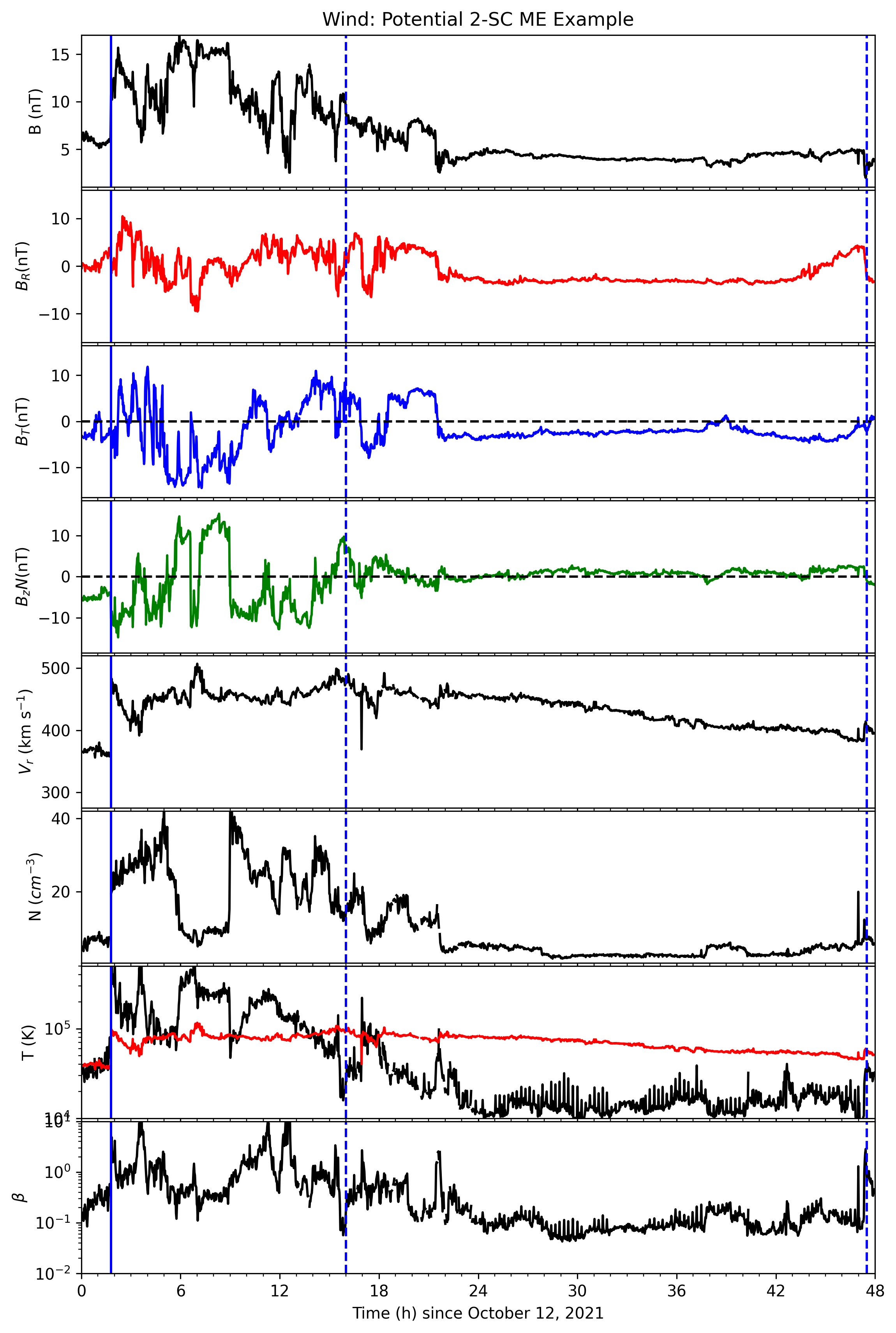}
	\includegraphics[width=0.4\textwidth]{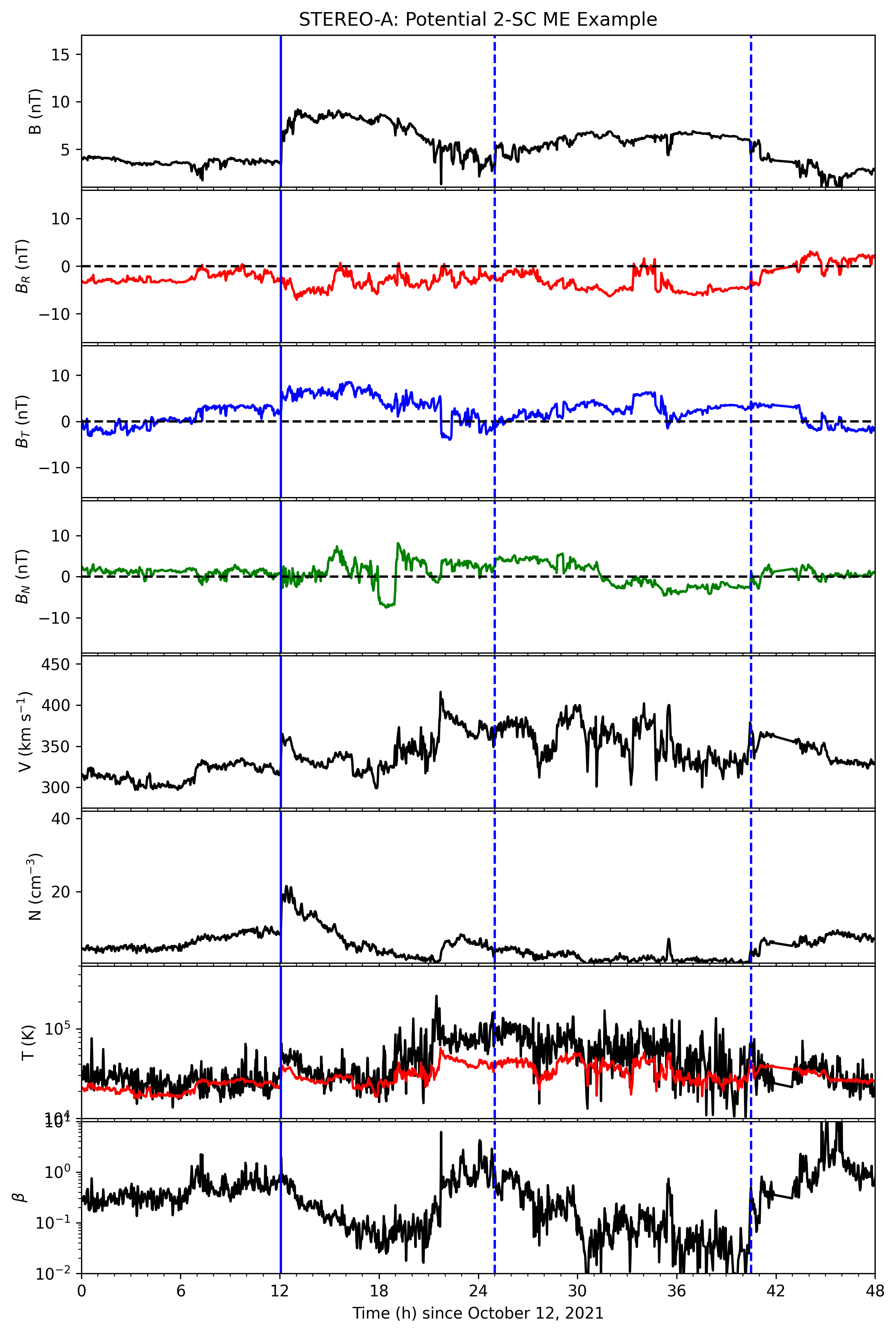}
\caption{{\it Wind} (left) and STEREO-A (right) measurements of a no detection of a ME for CME \#15 and a potential additional two-spacecraft measurement for a CME on 2021 October 12.  Same format as Figure~\ref{fig:two-SC}.} 
\label{fig:zero-SC}
\end{figure}

There is no CME listed either in HELIO4CAST or the database of \citet{Richardson:2010}. At {\it Wind}, there are plasma and magnetic field measurements that are consistent with a very weak ejecta, starting around 6~UT on March 28 and lasting about one day (see top left plot of Figure~\ref{fig:zero-SC}). There is a low density, lower than expected temperature, decreasing speed profile. However, the magnetic field is very weak, below 5~nT and the proton $\beta$ is not low indicating that this is not a magnetically dominated time period. At STEREO-A, there are clear indications of the passage of a stream interaction region (SIR), followed by a high speed solar wind stream. There is no period with lower than expected temperature or low proton $\beta$. It is not possible to prove that the signatures late on the March 27 and early on March 28 are not associated with a ME, but, if they do, they are very unusual.

Interaction with the high speed solar wind stream may have caused a deflection of the CME which resulted in the lack of any detection at {\it Wind} as well as a change of the morphology of the CME which makes the detection at STEREO-A more complex. Since the CME was initially propagating within about 5$^\circ$ of the Sun--STEREO-A line and 25$^\circ$ from the Sun--Earth line, this non-detection can only be understood by invoking the interaction with the SIR and high-speed stream. However, in light of the relatively low separations between STEREO-A and L1, this is still a surprising finding. 

\subsection{Effect of the CME Propagation Away from the Ecliptic}
Table~\ref{tab:CatI} shows in the eighth and ninth columns the latitudinal angles between the Sun-spacecraft line and the CME direction of propagation from the GCS reconstruction for {\it Wind} and STEREO-A. While both {\it Wind} and STEREO-A are in the ecliptic, they have different latitudes with respect to the solar equator, varying between $-6^\circ$ and $6^\circ$. The average latitudinal separation between the two spacecraft for these 21 CMEs is 2.5$^\circ$. 

We first note that there is no clear relationship between the lack or not of impact {\it in situ} and the latitudinal direction of the CME. For example, CME \#15 did not impact any spacecraft even though it was propagating within 5$^\circ$ of the ecliptic. On the other hand, CME \#21 impacted both STEREO-A and {\it Wind} even though it was $-35^\circ$ from the ecliptic. While strong latitudinal deflections have been reported in the low-to-middle corona \citep[]{Kilpua:2009b}, further latitudinal deflection in the upper corona has also been reported \citep[]{Byrne:2010}. Such deflections typically occur towards the ecliptic, so the number in Table~\ref{tab:CatI}, obtained from the GCS reconstruction using COR2 and C3 observations around 10~$R_\odot$ should be taken as an upper bound. 

Secondly, for sixteen of the 21 events, the difference in latitude between the Sun-spacecraft line and the CME direction of propagation is smaller than the difference in longitude. Some of the exceptions (CMEs \#8, \#11 and \#21) impacted at least one spacecraft {\it in situ}, so the impact of the latitudinal difference cannot be clearly determined. In fact, the current study focuses on the longitudinal size of the ME because STEREO-A and {\it Wind} are separated in longitude. An investigation of the ME latitudinal extent would be best performed using a much larger number of events, since it can only rely on single-spacecraft crossings. We discuss the potential effect of ME inclination in the discussion section (\ref{sec:newdicuss}).

Thirdly, it is in theory possible that the tilt between the ME and the ecliptic (where both STEREO-A and {\it Wind} are) results in additional misses. As shown in Table~\ref{tab:CatI}, there aren't any events where the ME propagates latitudinal in-between STEREO-A and {\it Wind}. However, for MEs with a significant inclination, such effects could still occur, {\it i.e.} the CME passes locally above one spacecraft and below the other. The current study is only able to determine the extent of the ME in the ecliptic, which is a combination of the cross-section shape, the ME leg-to-leg extent and the tilt of the ME with respect to the ecliptic. This is further discussed in the discussion section (\ref{sec:old_discuss}).

\section{{\it In Situ} Measurements of CMEs From November 2020 to August 2022}\label{sec:insitu}

\subsection{Methodology}
We perform a complementary study, where we start from {\it in situ} measurements at {\it Wind} and STEREO-A to ensure that there weren't additional two-spacecraft measurements of MEs during this time period which may not have been clearly associated with coronagraphic sources, either because they are associated with stealth CMEs (from both STEREO-A and LASCO points-of-view) or because interplanetary deflection, even if it is limited to a few degrees, resulted in a CME from beyond 30$^\circ$ from the Sun--spacecraft line to impact a given spacecraft.

We start from the list of CMEs at L1 from \citet{Richardson:2010} and the HELIO4CAST database for both CMEs at L1 and STEREO-A. We remove all events with ME duration shorter than 12 hours, as such events have been in the past described as small transients \citep[]{Cartwright:2010,Yu:2016} and may not always be of coronal origin. We also remove events with weak magnetic field, on average below $\sim$6~nT as these are significantly weaker than a typical ME near 1~au. We visually inspected all {\it in situ} measurements and ended up with 23 CMEs measured {\it in situ} by {\it Wind} and 17 by STEREO-A. We note that here MEs measured simultaneously at two spacecraft are double-counted. 

We found no additional clear two-spacecraft measurements outside of the four highlighted earlier. There is, however, one event, the 2021 October 12 ME, for which there might be multi-spacecraft measurements, which we briefly describe  below. 

\subsection{2021 October 12 ME}
There is a MC-1 in the database of \citet{Richardson:2010} at {\it Wind}, with a shock at 01:48 UT on October 12 and a ME from 16 UT on October 12 to the end of October 13 (see bottom panel of Figure~\ref{fig:zero-SC}). There is a clear FF shock and a dense, hot and magnetized sheath region followed by a colder-than-expected and relatively low $\beta$ ME. The magnetic field strength in the ME is on average 6~nT, $B_N \sim 0$~nT and there is only rotation in the first $\sim 5$ hours followed by steady magnetic field vectors. HELIO4CAST lists a very short ejecta ($\sim$ 4 hours) in the sheath region. Due to the steady nature of the magnetic field and the cold temperature, we follow the boundaries of \citet{Richardson:2010}. HELIO4CAST lists a CME at STEREO-A starting with a FF shock at 12:04 UT and a ME lasting 15.5 hours starting at 01 UT on October 13. While the magnetic field is also weak, there is a clearer rotation of the magnetic field vector with $B_N$ going from positive to negative. STEREO-A and {\it Wind} were separated by 38.6$^\circ$ in longitude. Based on the {\it in situ} measurements, it is impossible, in our opinion, to conclude whether or not this is the same event as there are not many similarities but the two events start within less than 12 hours from each other. The event at {\it Wind} is likely to be associated with the halo CME that started on 2021 October 9 around 7:12 UT, for which the GCS reconstruction shows a direction of propagation 3$^\circ$ west of the Sun--Earth line (41$^\circ$ from the Sun--STEREO-A line), a Cat II event. 

\subsection{Statistics}
As such, there are 35 or 36 individual CMEs measured either at {\it Wind}, at STEREO-A, or at both spacecraft, four of which have MEs clearly measured by both spacecraft and one is unclear. We do not consider events for which the shock is measured by one spacecraft and the ME by another as it is nearly impossible to confirm that they are associated with the same eruption. Overall, this indicates that 11--14\% of MEs measured {\it in situ} during this time period when STEREO-A and L1 were separated by 20--60$^\circ$ were measured by both spacecraft. This number is lower than the number of multi-spacecraft measurements based on remote observations ($\sim$ 20\%) because it includes (a) impacts from CMEs that were propagating west of the Sun--Earth line and impacted {\it Wind}, and (b) CMEs that were propagating east of the Sun--STEREO-A line and impacted STEREO-A. %We note that there were 21 Cat I CMEs propagating in-between the two spacecraft and 14 Cat II CMEs propagating 30$^\circ$ from the Sun--Earth line but beyond 30$^\circ$ from the Sun--STEREO-A line. Overall the number from {\it in situ} measurements should be about half that from remote observations. 

\section{Discussions}\label{sec:discussions}

\subsection{Additional Effects That May Influence the Results}\label{sec:newdicuss}
Under the flux rope paradigm for MEs, the angular extent in the ecliptic is a combination of the ME leg-to-leg width and the ME non-radial cross-section width. First, based on this paradigm, the likelihood of measuring the same ME at two spacecraft is expected to depend on the ME orientation: it should be high for MEs with low inclination and low for MEs with high inclination.
Assuming a circular cross-section, the ME typical radial size of 0.21~au corresponds to a non-radial cross-section angular width of 12$^\circ$. If the ME has an elliptical cross-section with a 3:1 ratio of the non-radial to radial widths, as found in \citet{Demoulin:2013} based on the distribution of impact parameters, ME would have an angular width of $\sim 37^\circ$. Under the model of \citet{Owens:2006a}, the ME angular width remains constant and should be $\sim$ 40--80$^\circ$. The leg-to-leg width is less well constrained but is typically assumed to be at least 60$^\circ$. 

We first note that most of the MEs plotted in this article are complex, often with multiple magnetic field vector rotations, and therefore determining the ME inclination is difficult. In addition, we do not rely on fitting techniques developed for single-spacecraft measurements, as such techniques can return results with large uncertainties \citep[]{AlHaddad:2013} and they rely on the highly twisted axisymmetric flux rope paradigm \citep[]{AlHaddad:2019b}, which we argue may not necessarily hold true in light of the presented measurements.  %For two events, CME \#12, measured at {\it Wind} and STEREO-A and CME \#9 measured at {\it Wind}, the magnetic field measurements are consistent with that of highly inclined MEs. Overall, 
Based on our limited sample, we do not find any indication that the ME magnetic field orientation has an impact on the ME angular width (or the likelihood of measuring the same ME at two different locations). 
Low-inclined MEs should be easily measured by two spacecraft for separations of 45--60$^\circ$ if the expectation about the leg-to-leg width is correct, yet we do not see any indication that more low-inclined MEs were measured at large separations. This puts into question the flux rope paradigm. 

CME-CME interaction has been reported to sometimes result in longitudinal deflection \citep[]{Lugaz:2012b} and in compression of the ME in the radial direction \citep[e.g., see][]{Lugaz:2005b}, and it is likely to also result in changes in the CME angular width. Deflections in particular may result in CMEs having a different direction in the heliosphere as that obtained from GCS.
Additionally, CMEs often interact with SIRs as they propagate \citep[]{Farrugia:2011, Prise:2015, Winslow:2016,Heinemann:2019}. This can result in significant differences in the ME properties for events measured near simultaneously by two spacecraft as is presented here \citep[]{Winslow:2021,Lugaz:2022}. Deflection of CMEs in the heliosphere due to their interaction with the structured solar wind have also been hypothesized \citep[]{Wang:2004} and may influence the presented results.

Overall, the current study focuses on the impact of MEs at the two spacecraft to deduce the angular extent in the ecliptic plane, not the differences in the properties of the MEs at the two spacecraft. For this reason, we did not perform an in-depth study of the interplanetary conditions into which the CMEs propagate. Therefore, CME-CME and CME-SIR interactions may influence the presented results.

\subsection{CME Angular Size in the Ecliptic}\label{sec:old_discuss}

11--14\% of MEs measured {\it in situ} during the time period from November 2020 to August 2022 when STEREO-A and L1 were separated by 20--60$^\circ$ in longitude were measured {\it in situ} by both spacecraft. This is a slightly lower number than that obtained by \citet{Good:2016}, which found 7 out of 39 events (18\%) for two-spacecraft separations of 30--45$^\circ$. However,  they found a much higher number (19 out 39, so 48\%) for separations of 15--30$^\circ$. Their studies had two main limitations: the two spacecraft were at different distances and one of the two spacecraft typically did not have plasma measurements. We therefore consider that the number obtained by \citet{Good:2016} represents an upper bound for the proportion of MEs that may be measured {\it in situ} by two spacecraft near 1~au. 

One way to analyze the results of the remote-sensing investigation is as follows: for an average separation between the central direction of a CME as obtained from the GCS reconstruction and the Sun--spacecraft line of 18.9$^\circ$ (for our 21 events), only two thirds of the CMEs impact a spacecraft. This would indicate that the median angular size in the ecliptic of MEs is close to 40$^\circ$. However, this does not take into consideration the fact that there were two spacecraft potentially taking {\it in situ} measurements. For the 21 Cat I CMEs, the average STEREO-A--{\it Wind} separation is only 37.6$^\circ$, however there are only four multi-spacecraft ME measurements. This indicates a typical angular size in the ecliptic significantly lower than 30$^\circ$. Out of these 21 events, there were nine CMEs that are found to propagate almost exactly in-between the two-spacecraft (within $\pm 5^\circ$ of the bisector) with an average two-spacecraft separation of 40.2$^\circ \pm 13.9^\circ$. Two of these CMEs impacted both spacecraft, six impacted one spacecraft and one impacted none (for the largest separation of 57$^\circ$). This indicates that the ME angular size in the ecliptic might be as small as 20$^\circ$. 

Overall, we consider that this analysis indicates that the ME angular size in the ecliptic is likely to be 20--30$^\circ$, which corresponds to a size of $\sim 0.35-0.5$~au, or 2 $\pm 0.4$ times larger than the radial size of MEs. As we do not see any influence of the ME orientation, we consider that a model where the leg-to-leg and non-radial cross-section width are more or less equal is the most likely interpretation of the data. This would be closer to the geometry of an ice cream cone model or spheromak rather than a flux rope model for the ME. Such models, in recent years, are often considered as less realistic than flux rope models of CMEs \citep[see, for example:][]{Luhmann:2020}. The ME is however not exactly spherical but corresponds to two scoops of ice cream (aspect ratio of $\sim$ 2:1). 

The few multi-spacecraft measurements of MEs that we found indicate the magnetic field is not coherent between the two spacecraft, even when two spacecraft are able to measure the ME simultaneously. This is because the magnetic field and plasma signatures differ significantly between the two spacecraft (see Figures~\ref{fig:two-SC} and \ref{fig:one-SC}). In fact in most of these four cases, without the knowledge that one and only one CME was propagating in-between the two spacecraft, it is unlikely that one could be sure that the same event is measured {\it in situ} by STEREO-A and {\it Wind}. This is particularly true for event \#5 and \#12. None of these events present MC-like signatures at both spacecraft (CMEs \#5 and \#12 are MC-like at STEREO-A only) and performing an analysis of the correlation between the magnetic field measurements as done in \citet{Lugaz:2018} would be meaningless since the measurements are too different. 
Overall, this indicates that past estimates of the ME coherence of $\sim 20^\circ$ \citep[]{Lugaz:2018,Scolini:2023} may be an upper bound for the true coherence length. We note that, based on the analysis of \citet{Owens:2017}, a ME with an angular width of $20^\circ$ should be coherent if Alfv\'en waves are the carrier of information across the ME. 

\subsection{Consequences for Future Mission Concepts}
We first mention that missions at the Sun--Earth Lagrangian L4 or L5 points may provide critical information from remote observations of CMEs but will not provide {\it in situ} multi-spacecraft ME observations, except in extremely rare cases. This is because we did not measure any CMEs for separations beyond 55.6$^\circ$ as presented in \citet{Lugaz:2022}. In addition, s shown in \citet{Bailey:2020}, measurements of the magnetic field (especially $B_N$) of corotating streams are poorly correlated between L1 and L5.

Numerous multi-spacecraft {\it in situ} mission concepts have been recently proposed to investigate CMEs, shocks and solar energetic particles \citep[e.g., see white papers to the 2023 Decadal Survey in][]{Allen:2022,Lugaz:2023a,Nykyri:2023,Akhavan-Tafti:2023}, some of them based on earlier concepts \citep[]{StCyr:2000b,Szabo:2005}. The inter-spacecraft separations have been proposed to vary between $60^\circ$--$90^\circ$ for the HELIX mission concept and less than 1$^\circ$ for the SWIFT concept of \citet{Akhavan-Tafti:2023}. We first note that the separations needed to investigate ME magnetic field and coherence may differ significantly from the separations needed to investigate SEP spread due to field line meandering or cross-field diffusion, and from the separations needed to investigate the global structure of shocks. Based on the current study, separations of more than 20$^\circ$ are not appropriate to investigate ME magnetic field as multi-spacecraft measurements are rare at these separations. Also, when they occur, multi-spacecraft measurements at separations of more than 20$^\circ$  do not provide useful information to create more complex models of the ME magnetic field. In fact, longitudinal separations well below 20$^\circ$ (for example $\sim$ 10$^\circ$) appear to be more appropriate.

As our study covers 22 months during the ascending phase of the solar cycle, when there were about 20 CMEs measured {\it in situ} at L1, it may also provide an estimate for the number of multi-spacecraft measurements that may occur during the cruise phase of a mission to the Sun--Earth L4 or L5 points, which is typically designed to take 2--3 years. It is likely that no more than 25\% of CMEs will be measured {\it in situ} at L1 and the cruising spacecraft. Such numbers should be used to further develop mission concepts and quantifying level-1 requirements. 

\section{Summary and Conclusions}\label{sec:conclusions}
We have investigated the angular width of MEs by studying all CMEs that were observed remotely to propagate between the Sun--Earth and Sun--STEREO-A lines from November 2020 to August 2022, when the STEREO-A--Sun--Earth angle was between 20$^\circ$ and 60$^\circ$. We identified 21 such events. The key result is that only four of these 21 CMEs were measured by both STEREO-A and {\it Wind}, for separations of 55.6$^\circ$, 51$^\circ$, 34.7$^\circ$ and 21$^\circ$. 10 others were measured by only one spacecraft and the last seven by none. In addition, for the four CMEs measured simultaneously by {\it Wind} and STEREO-A, the plasma and magnetic field measurements at the two spacecraft differ significantly, which highlights that the coherence length is smaller than the two-spacecraft separation. 

We complemented the study by looking at all MEs measured {\it in situ} by {\it Wind} or STEREO-A during the same time period and identifying any additional two-spacecraft measurements. We found one potential additional two-spacecraft measurement possibly associated with a CME propagating 3$^\circ$ of the Sun--Earth line (and 41$^\circ$ of the Sun--STEREO-A line). 

We draw the following conclusions from this study: (i) ME angular width in the ecliptic is likely to be only 20--30$^\circ$ for most events, extending to $\sim$ 50-55$^\circ$ for the largest events; (ii) there is little indication that low-inclined CMEs have a wider width than highly inclined ones; (iii) therefore, CME models need to account for an elliptical cross-section with a relatively small eccentricity (2:1 ratio of major to minor axes) and similar angular sizes from leg-to-leg as compared to the maximum size of the cross-section. These results also highlight that the small number of multi-spacecraft measurements of CMEs during the first two years of the mission (2007--2008) may have been a normal occurrence, not only due to the solar minimum conditions. \citet{Kilpua:2011} found four clear multi-spacecraft events during this time period, but with events like the 2007 May 23 ME measured {\it in situ} by only one of the two STEREO spacecraft while they were separated by 9$^\circ$. STEREO-A has now passed in front of the Sun--Earth line in August 2023 and the time period from August 2022 to July 2024 shall provide numerous multi-spacecraft measurements of CMEs for two-spacecraft separations of $\pm 20^\circ$ during or close to solar maximum conditions. The time period from mid-2024 to May 2026 would provide a complementary study to the one presented here, for the early part of the descending phase of the solar cycle, if STEREO-A continues to take data in the next three years.  

\appendix

\begin{figure}[h!]
	\centering
        \includegraphics[width=\textwidth]{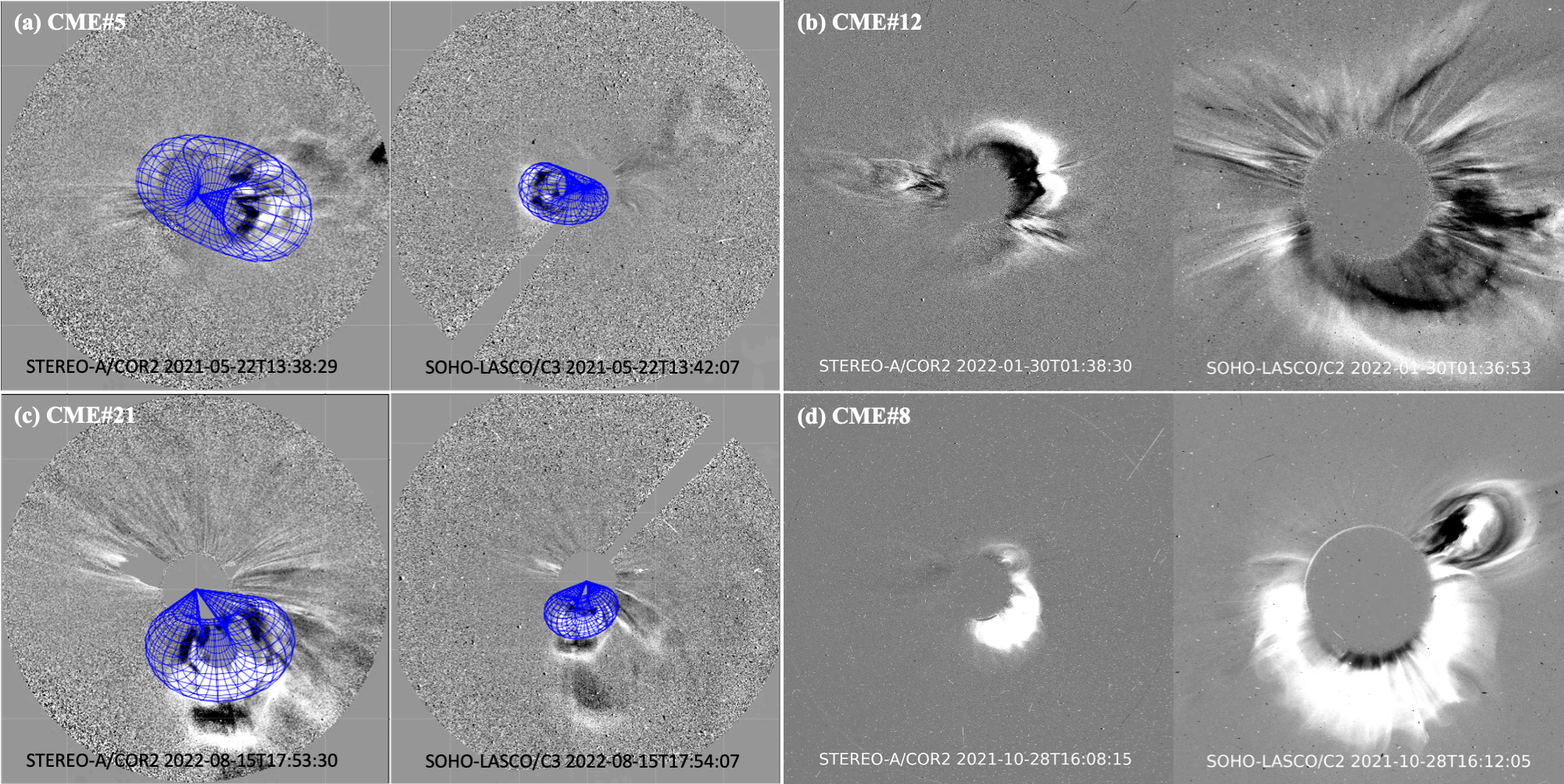}
\caption{Remote-sensing observations from STEREO-A/COR2 (left) and SOHO/LASCO (right) of the CMEs discussed in Figures~\ref{fig:two-SC} and \ref{fig:one-SC}, from top left to bottom right, CMEs \#5, \#12, \#21 (all impacting both spacecraft) and CME \#8 (impacting STEREO-A only). When the CME front or direction is not clear, the GCS reconstruction is also shown in blue overlay.} %When the CME front is not clear, .} 
\label{fig:appendixOne}
\end{figure}

\begin{figure}[h!]
	\centering
        \includegraphics[width=\textwidth]{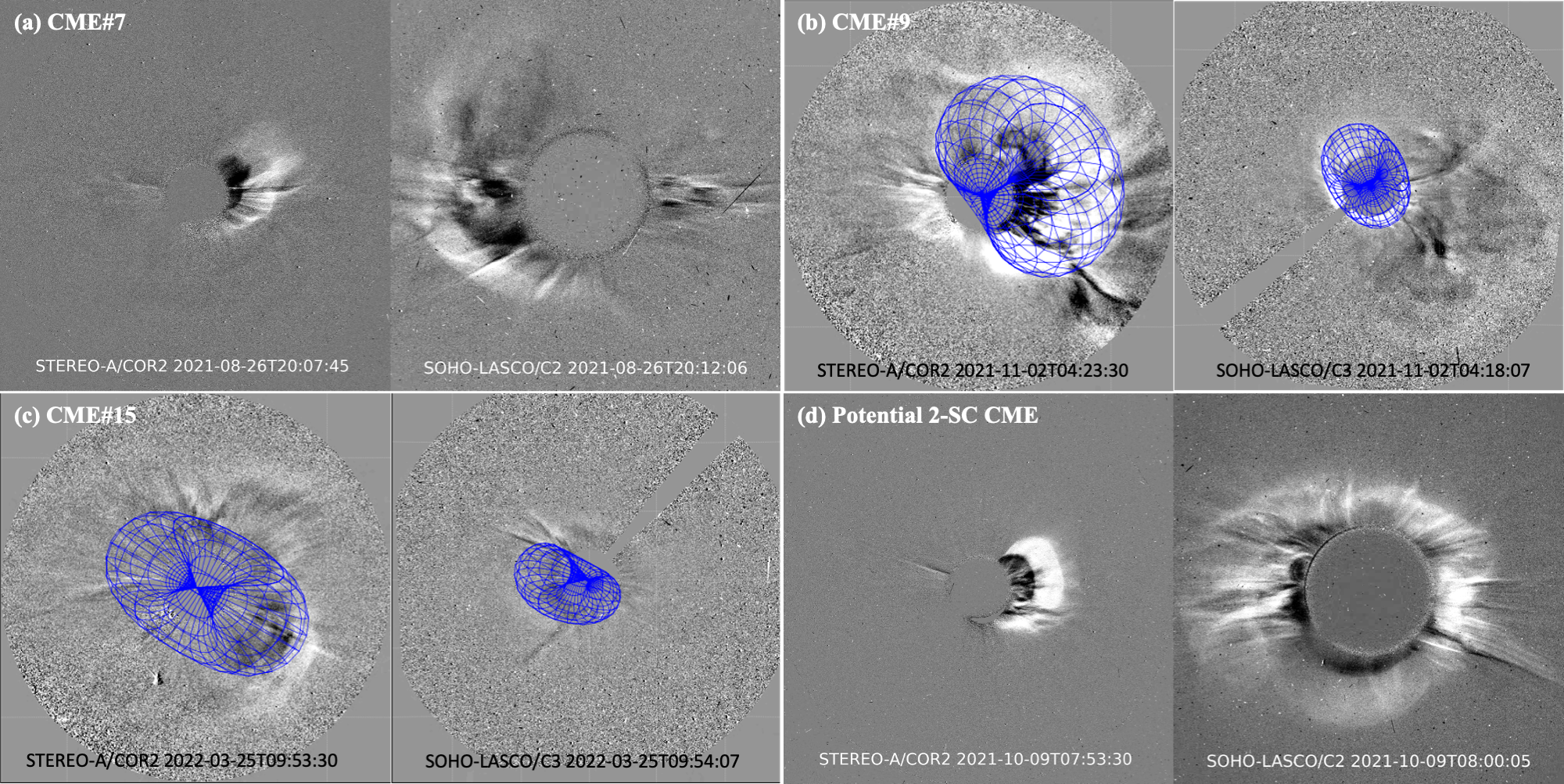}
\caption{Same as Figure~\ref{fig:appendixOne} for the CMEs discussed in Figures~\ref{fig:onetrue-SC} and \ref{fig:zero-SC}, from top left to bottom right, CME \#7 (impacting STEREO-A only), CME \#9  (impacting {\it Wind} only), CME \#15 (no impact) and the additional CME that may have two-spacecraft measurements.}  
\label{fig:appendixTwo}
\end{figure}

In Table~\ref{tab:appendix}, we list the properties of the Cat II, III and IV CMEs. In Figures~\ref{fig:appendixOne} and \ref{fig:appendixTwo}, we show remote images of the CMEs analyzed in this study.

\begin{table}[!ht]
    \centering
    \caption{List of Cat. II, Cat III and Cat IV CMEs. Potential impact for Cat. III and Cat. IV CMEs was not investigated. Back refers to back-sided events as seen from L1.}
    \begin{tabular}{|c|c|c|c|c|c|}
    \hline
        {\bf $\Delta \theta$ ($^\circ$)} & {\bf Date} & {\bf Time (UT)} & {\bf V (km\,s$^{-1}$)} & {\bf $\phi_{L1}$} & {\bf Impact} \\ \hline
        \multicolumn{6}{|l|}{\bf Cat II} \\ \hline
        57.5 & 7-Dec-20 & 16:24 & 1407 & 13 & None \\ \hline
        53.1 & 22-Apr-21 & 05:48 & 355 & 5 & None \\ \hline
        52 & 9-May-21 & 12:00 & 266 & 4 & Wind \\ \hline
        38.6 & 9-Oct-21 & 07:12 & 712 & 3 & Unclear \\ \hline
        34 & 6-Feb-22 & 14:00 & 334 & 3 & Wind \\ \hline
        33.8 & 7-Mar-22 & 00:12 & 241 & 26 & None \\ \hline
        33.7 & 10-Mar-22 & 18:48 & 742 & 8 & Wind \\ \hline
        32.7 & 3-Apr-22 & 16:48 & 502 & 18 & None \\ \hline
        32.9 & 28-Mar-22 & 20:24 & 905 & 12 & None \\ \hline
        32.8 & 30-Mar-22 & 18:00 & 641 & 24 & None \\ \hline
        32.2 & 11-Apr-22 & 05:48 & 940 & 0 & Wind \\ \hline
        24.4 & 15-Jul-22 & 15:48 & 320 & 19 & Wind \\ \hline
        21 & 17-Aug-22 & 14:36 & 901 & 21 & None \\ \hline
        21 & 14-Aug-22 & 12:48 & 481 & 18 & None \\ \hline
        \multicolumn{6}{|l|}{\bf Cat III} \\ \hline
        52 & 9-May-21 & 15:23 & 603 & 61 & ~ \\ \hline
        45.2 & 24-Jul-21 & 00:36 & 587 & 57 & ~ \\ \hline
        47 & 31-Oct-21 & 08:12 & 366 & 47 & ~ \\ \hline
        27.5 & 13-Jun-22 & 03:12 & 1150 & 48 & ~ \\ \hline
        \multicolumn{6}{|l|}{\bf Cat IV} \\ \hline
        57.8 & 29-Nov-20 & 13:25 & 2077 & 96 & ~ \\ \hline
        50.5 & 28-May-21 & 23:12 & 971 & 56 & ~ \\ \hline
        38.7 & 7-Oct-21 & 02:00 & 299 & 90 & ~ \\ \hline
        34.4 & 11-Feb-22 & 00:48 & 448 & 143 & ~ \\ \hline
        31 & 1-May-22 & 08:36 & 700 & 106 & ~ \\ \hline
        30.5 & 8-May-22 & 04:24 & 263 & Back & ~ \\ \hline
        28.9 & 28-May-22 & 15:12 & 568 & Back & ~ \\ \hline
        27.4 & 14-Jun-22 & 01:25 & 662 & Back & ~ \\ \hline
        26.2 & 26-Jun-22 & 03:36 & 980 & Back & ~ \\ \hline
        23.6 & 23-Jul-22 & 19:00 & 1026 & Back & ~ \\ \hline
        22.9 & 30-Jul-22 & 02:24 & 609 & Back & ~ \\ \hline
        20.3 & 26-Aug-22 & 10:36 & 553 & 62 & ~\\ \hline
    \end{tabular}
    \label{tab:appendix}
\end{table}

\begin{acknowledgments}

The dataset used in this study are from STEREO-A/PLASTIC and IMPACT and {\it Wind} MFI and SWE as retrieved from NASA CDAWeb servers.

The HELIO4CAST website is \url{https://helioforecast.space/icmecat} with the associated doi: \\
10.6084/m9.figshare.6356420

All UNH authors acknowledge grant 80NSSC20K0431.  N.~L. acknowledges additional support from 80NSSC20K0700. B.~Z. and C.~S. acknowledge  additional support from 80NSSC23K1057. B.~Z. acknowledges additional support from AGS-2301382. N.~A. and F.~R. acknowledge grants 80NSSC22K0349, 80NSSC21K0463  and AGS1954983. C.~J.~F. acknowledges support from grants 80NSSC21K0463, and Winds's grant 80NSSC19K1293. E.~D. and C.~M. acknowledge funding by the European Union (ERC, HELIO4CAST, 101042188). Views and opinions expressed are however those of the author(s) only and do not necessarily reflect those of the European Union or the European Research Council Executive Agency. Neither the European Union nor the granting authority can be held responsible for them.

\end{acknowledgments}


\begin{thebibliography}{}
\expandafter\ifx\csname natexlab\endcsname\relax\def\natexlab#1{#1}\fi
\providecommand{\url}[1]{\href{#1}{#1}}
\providecommand{\dodoi}[1]{doi:~\href{http://doi.org/#1}{\nolinkurl{#1}}}
\providecommand{\doeprint}[1]{\href{http://ascl.net/#1}{\nolinkurl{http://ascl.net/#1}}}
\providecommand{\doarXiv}[1]{\href{https://arxiv.org/abs/#1}{\nolinkurl{https://arxiv.org/abs/#1}}}

\bibitem[{{Akhavan-Tafti} {et~al.}(2023){Akhavan-Tafti}, {Johnson}, {Sood},
  {Slavin}, {Pulkkinen}, {Lepri}, {Kilpua}, {Fontaine}, {Szabo}, {Wilson},
  {Le}, {Atilaw}, {Ala-Lahti}, {Soni}, {Biesecker}, {Jian}, \&
  {Lario}}]{Akhavan-Tafti:2023}
{Akhavan-Tafti}, M., {Johnson}, L., {Sood}, R., {et~al.} 2023, Frontiers in
  Astronomy and Space Sciences, 10, 1185603, \dodoi{10.3389/fspas.2023.1185603}

\bibitem[{{Al-Haddad} {et~al.}(2022){Al-Haddad}, {Galvin}, {Lugaz}, {Farrugia},
  \& {Yu}}]{AlHaddad:2022}
{Al-Haddad}, N., {Galvin}, A.~B., {Lugaz}, N., {Farrugia}, C.~J., \& {Yu}, W.
  2022, Astrophys. J., 927, 68, \dodoi{10.3847/1538-4357/ac32e1}

\bibitem[{{Al-Haddad} {et~al.}(2019){Al-Haddad}, {Poedts}, {Roussev},
  {Farrugia}, {Yu}, \& {Lugaz}}]{AlHaddad:2019b}
{Al-Haddad}, N., {Poedts}, S., {Roussev}, I., {et~al.} 2019, Astrophys. J.,
  870, 100, \dodoi{10.3847/1538-4357/aaf38d}

\bibitem[{{Al-Haddad} {et~al.}(2013){Al-Haddad}, {Nieves-Chinchilla},
  {M{\"o}stl}, {Hidalgo}, {Marubashi}, {Savani}, {Roussev}, {Poedts}, \&
  {Farrugia}}]{AlHaddad:2013}
{Al-Haddad}, N., {Nieves-Chinchilla}, T., {M{\"o}stl}, C., {et~al.} 2013, Sol.
  Phys., 284, 129, \dodoi{10.1007/s11207-013-0244-5}

\bibitem[{{Allen} {et~al.}(2022){Allen}, {Smith}, {Anderson}, {Borovsky}, {Ho},
  {Jian}, {Krucker}, {Lepri}, {Li}, {Livi}, {Lugaz}, {Malaspina}, {Maruca},
  {Mostafavi}, {Raines}, {Verscharen}, {Vievering}, {Vines}, {Whittlesey},
  {Wilson}, \& {Wimmer-Schweingruber}}]{Allen:2022}
{Allen}, R.~C., {Smith}, E.~J., {Anderson}, B.~J., {et~al.} 2022, Frontiers in
  Astronomy and Space Sciences, 9, 1002273, \dodoi{10.3389/fspas.2022.1002273}

\bibitem[{{Bailey} {et~al.}(2020){Bailey}, {M{\"o}stl}, {Reiss}, {Weiss},
  {Amerstorfer}, {Amerstorfer}, {Hinterreiter}, {Magnes}, \&
  {Leonhardt}}]{Bailey:2020}
{Bailey}, R.~L., {M{\"o}stl}, C., {Reiss}, M.~A., {et~al.} 2020, Space Weather,
  18, e02424, \dodoi{10.1029/2019SW002424}

\bibitem[{{Brueckner} {et~al.}(1995){Brueckner}, {Howard}, {Koomen},
  {Korendyke}, {Michels}, {Moses}, {Socker}, {Dere}, {Lamy}, {Llebaria},
  {Bout}, {Schwenn}, {Simnett}, {Bedford}, \& {Eyles}}]{Brueckner:1995}
{Brueckner}, G.~E., {Howard}, R.~A., {Koomen}, M.~J., {et~al.} 1995, Sol.
  Phys., 162, 357, \dodoi{10.1007/BF00733434}

\bibitem[{{Burlaga} {et~al.}(1981){Burlaga}, {Sittler}, {Mariani}, \&
  {Schwenn}}]{Burlaga:1981}
{Burlaga}, L., {Sittler}, E., {Mariani}, F., \& {Schwenn}, R. 1981, J. Geophys.
  Res., 86, 6673, \dodoi{10.1029/JA086iA08p06673}

\bibitem[{{Byrne} {et~al.}(2010){Byrne}, {Maloney}, {McAteer}, {Refojo}, \&
  {Gallagher}}]{Byrne:2010}
{Byrne}, J.~P., {Maloney}, S.~A., {McAteer}, R.~T.~J., {Refojo}, J.~M., \&
  {Gallagher}, P.~T. 2010, Nature Communications, 1, \dodoi{10.1038/ncomms1077}

\bibitem[{{Cane} {et~al.}(1998){Cane}, {Richardson}, \& {St Cyr}}]{Cane:1998}
{Cane}, H.~V., {Richardson}, I.~G., \& {St Cyr}, O.~C. 1998, Geophys. Res.
  Lett., 25, 2517, \dodoi{10.1029/98GL00494}

\bibitem[{{Cartwright} \& {Moldwin}(2010)}]{Cartwright:2010}
{Cartwright}, M.~L., \& {Moldwin}, M.~B. 2010, J. Geophys. Res., 115, 8102,
  \dodoi{10.1029/2009JA014271}

\bibitem[{{Davies} {et~al.}(2012){Davies}, {Harrison}, {Perry}, {M{\"o}stl},
  {Lugaz}, {Rollett}, {Davis}, {Crothers}, {Temmer}, {Eyles}, \&
  {Savani}}]{Davies:2012}
{Davies}, J.~A., {Harrison}, R.~A., {Perry}, C.~H., {et~al.} 2012, Astrophys.
  J., 750, 23, \dodoi{10.1088/0004-637X/750/1/23}

\bibitem[{{D{\'e}moulin} {et~al.}(2013){D{\'e}moulin}, {Dasso}, \&
  {Janvier}}]{Demoulin:2013}
{D{\'e}moulin}, P., {Dasso}, S., \& {Janvier}, M. 2013, Astron. Astrophys.,
  550, A3

\bibitem[{{D{\'e}moulin} {et~al.}(2016){D{\'e}moulin}, {Janvier},
  {Mas{\'{\i}}as-Meza}, \& {Dasso}}]{Demoulin:2016}
{D{\'e}moulin}, P., {Janvier}, M., {Mas{\'{\i}}as-Meza}, J.~J., \& {Dasso}, S.
  2016, Astron. Astrophys., 595, A19, \dodoi{10.1051/0004-6361/201628164}

\bibitem[{{Farrugia} {et~al.}(2011){Farrugia}, {Berdichevsky}, {M{\"o}stl},
  {Galvin}, {Leitner}, {Popecki}, {Simunac}, {Opitz}, {Lavraud}, {Ogilvie},
  {Veronig}, {Temmer}, {Luhmann}, \& {Sauvaud}}]{Farrugia:2011}
{Farrugia}, C.~J., {Berdichevsky}, D.~B., {M{\"o}stl}, C., {et~al.} 2011, J.
  Atmos. Solar-Terr. Phys., 73, 1254, \dodoi{10.1016/j.jastp.2010.09.011}

\bibitem[{{Fox} {et~al.}(2016){Fox}, {Velli}, {Bale}, {Decker}, {Driesman},
  {Howard}, {Kasper}, {Kinnison}, {Kusterer}, {Lario}, {Lockwood}, {McComas},
  {Raouafi}, \& {Szabo}}]{Fox:2016}
{Fox}, N.~J., {Velli}, M.~C., {Bale}, S.~D., {et~al.} 2016, Space Sci. Rev.,
  204, 7, \dodoi{10.1007/s11214-015-0211-6}

\bibitem[{{Galvin} {et~al.}(2008){Galvin}, {Kistler}, {Popecki}, {Farrugia},
  {Simunac}, {Ellis}, {M{\"o}bius}, {Lee}, {Boehm}, {Carroll}, {Crawshaw},
  {Conti}, {Demaine}, {Ellis}, {Gaidos}, {Googins}, {Granoff}, {Gustafson},
  {Heirtzler}, {King}, {Knauss}, {Levasseur}, {Longworth}, {Singer}, {Turco},
  {Vachon}, {Vosbury}, {Widholm}, {Blush}, {Karrer}, {Bochsler}, {Daoudi},
  {Etter}, {Fischer}, {Jost}, {Opitz}, {Sigrist}, {Wurz}, {Klecker}, {Ertl},
  {Seidenschwang}, {Wimmer-Schweingruber}, {Koeten}, {Thompson}, \&
  {Steinfeld}}]{Galvin:2008}
{Galvin}, A.~B., {Kistler}, L.~M., {Popecki}, M.~A., {et~al.} 2008, Space Sci.
  Rev., 136, 437, \dodoi{10.1007/s11214-007-9296-x}

\bibitem[{{Good} \& {Forsyth}(2016)}]{Good:2016}
{Good}, S.~W., \& {Forsyth}, R.~J. 2016, Sol. Phys., 291, 239,
  \dodoi{10.1007/s11207-015-0828-3}

\bibitem[{{Gopalswamy} {et~al.}(2015){Gopalswamy}, {Yashiro}, {Xie}, {Akiyama},
  \& {M{\"a}kel{\"a}}}]{Gopalswamy:2015}
{Gopalswamy}, N., {Yashiro}, S., {Xie}, H., {Akiyama}, S., \& {M{\"a}kel{\"a}},
  P. 2015, J. Geophys. Res., 120, 9221, \dodoi{10.1002/2015JA021446}

\bibitem[{{Heinemann} {et~al.}(2019){Heinemann}, {Temmer}, {Farrugia},
  {Dissauer}, {Kay}, {Wiegelmann}, {Dumbovi{\'c}}, {Veronig}, {Podladchikova},
  {Hofmeister}, {Lugaz}, \& {Carcaboso}}]{Heinemann:2019}
{Heinemann}, S.~G., {Temmer}, M., {Farrugia}, C.~J., {et~al.} 2019, Sol. Phys.,
  294, 121, \dodoi{10.1007/s11207-019-1515-6}

\bibitem[{{Howard} {et~al.}(2008){Howard}, {Moses}, {Vourlidas}, {Newmark},
  {Socker}, {Plunkett}, {Korendyke}, {Cook}, {Hurley}, {Davila}, {Thompson},
  {St Cyr}, {Mentzell}, {Mehalick}, {Lemen}, {Wuelser}, {Duncan}, {Tarbell},
  {Wolfson}, {Moore}, {Harrison}, {Waltham}, {Lang}, {Davis}, {Eyles},
  {Mapson-Menard}, {Simnett}, {Halain}, {Defise}, {Mazy}, {Rochus}, {Mercier},
  {Ravet}, {Delmotte}, {Auchere}, {Delaboudiniere}, {Bothmer}, {Deutsch},
  {Wang}, {Rich}, {Cooper}, {Stephens}, {Maahs}, {Baugh}, {McMullin}, \&
  {Carter}}]{Howard:2008}
{Howard}, R.~A., {Moses}, J.~D., {Vourlidas}, A., {et~al.} 2008, Space Sci.
  Rev., 136, 67, \dodoi{10.1007/s11214-008-9341-4}

\bibitem[{{Howard} \& {DeForest}(2012)}]{THoward:2012}
{Howard}, T.~A., \& {DeForest}, C.~E. 2012, Astrophys. J., 746, 64,
  \dodoi{10.1088/0004-637X/746/1/64}

\bibitem[{{Hundhausen}(1993)}]{Hundhausen:1993}
{Hundhausen}, A.~J. 1993, J. Geophys. Res., 98, 13177

\bibitem[{{Janvier} {et~al.}(2015){Janvier}, {Dasso}, {D{\'e}moulin},
  {Mas{\'{\i}}as-Meza}, \& {Lugaz}}]{Janvier:2015}
{Janvier}, M., {Dasso}, S., {D{\'e}moulin}, P., {Mas{\'{\i}}as-Meza}, J.~J., \&
  {Lugaz}, N. 2015, Journal of Geophysical Research (Space Physics), 120, 3328,
  \dodoi{10.1002/2014JA020836}

\bibitem[{{Janvier} {et~al.}(2014){Janvier}, {D{\'e}moulin}, \&
  {Dasso}}]{Janvier:2014a}
{Janvier}, M., {D{\'e}moulin}, P., \& {Dasso}, S. 2014, Astron. Astrophys.,
  565, A99, \dodoi{10.1051/0004-6361/201423450}

\bibitem[{{Jian} {et~al.}(2018){Jian}, {Russell}, {Luhmann}, \&
  {Galvin}}]{Jian:2018}
{Jian}, L.~K., {Russell}, C.~T., {Luhmann}, J.~G., \& {Galvin}, A.~B. 2018,
  Astrophys. J., 855, 114, \dodoi{10.3847/1538-4357/aab189}

\bibitem[{{Kilpua} {et~al.}(2017){Kilpua}, {Koskinen}, \&
  {Pulkkinen}}]{Kilpua:2017}
{Kilpua}, E., {Koskinen}, H.~E.~J., \& {Pulkkinen}, T.~I. 2017, Liv. Rev. Sol.
  Phys., 14, 5, \dodoi{10.1007/s41116-017-0009-6}

\bibitem[{{Kilpua} {et~al.}(2011){Kilpua}, {Lee}, {Luhmann}, \&
  {Li}}]{Kilpua:2011}
{Kilpua}, E.~K.~J., {Lee}, C.~O., {Luhmann}, J.~G., \& {Li}, Y. 2011, Annales
  Geophysicae, 29, 1455, \dodoi{10.5194/angeo-29-1455-2011}

\bibitem[{{Kilpua} {et~al.}(2009){Kilpua}, {Pomoell}, {Vourlidas}, {Vainio},
  {Luhmann}, {Li}, {Schroeder}, {Galvin}, \& {Simunac}}]{Kilpua:2009b}
{Kilpua}, E.~K.~J., {Pomoell}, J., {Vourlidas}, A., {et~al.} 2009, Annales
  Geophysicae, 27, 4491, \dodoi{10.5194/angeo-27-4491-2009}

\bibitem[{{Lemen} {et~al.}(2012){Lemen}, {Title}, {Akin}, {Boerner}, {Chou},
  {Drake}, {Duncan}, {Edwards}, {Friedlaender}, {Heyman}, {Hurlburt}, {Katz},
  {Kushner}, {Levay}, {Lindgren}, {Mathur}, {McFeaters}, {Mitchell}, {Rehse},
  {Schrijver}, {Springer}, {Stern}, {Tarbell}, {Wuelser}, {Wolfson}, {Yanari},
  {Bookbinder}, {Cheimets}, {Caldwell}, {Deluca}, {Gates}, {Golub}, {Park},
  {Podgorski}, {Bush}, {Scherrer}, {Gummin}, {Smith}, {Auker}, {Jerram},
  {Pool}, {Soufli}, {Windt}, {Beardsley}, {Clapp}, {Lang}, \&
  {Waltham}}]{Lemen:2012}
{Lemen}, J.~R., {Title}, A.~M., {Akin}, D.~J., {et~al.} 2012, Sol. Phys., 275,
  17, \dodoi{10.1007/s11207-011-9776-8}

\bibitem[{{Lepping} {et~al.}(1995){Lepping}, {Acuna}, {Burlaga}, \&
  M.}]{Lepping:1995}
{Lepping}, R.~P., {Acuna}, M.~H., {Burlaga}, L.~F., \& M., F.~W. 1995, Space
  Sci. Rev., 71, 207

\bibitem[{{Lepping} {et~al.}(2005){Lepping}, {Wu}, \&
  {Berdichevsky}}]{Lepping:2005}
{Lepping}, R.~P., {Wu}, C.-C., \& {Berdichevsky}, D.~B. 2005, Annales
  Geophysicae, 23, 2687, \dodoi{10.5194/angeo-23-2687-2005}

\bibitem[{{Lin} {et~al.}(1995){Lin}, {Anderson}, {Ashford}, {Carlson},
  {Curtis}, {Ergun}, {Larson}, {McFadden}, {McCarthy}, {Parks}, {R{\`e}me},
  {Bosqued}, {Coutelier}, {Cotin}, {D'Uston}, {Wenzel}, {Sanderson}, {Henrion},
  {Ronnet}, \& {Paschmann}}]{Lin:1995}
{Lin}, R.~P., {Anderson}, K.~A., {Ashford}, S., {et~al.} 1995, Space Sci. Rev.,
  71, 125, \dodoi{10.1007/BF00751328}

\bibitem[{{Lopez}(1987)}]{Lopez:1987}
{Lopez}, R.~E. 1987, J. Geophys. Res., 92, 11189,
  \dodoi{10.1029/JA092iA10p11189}

\bibitem[{{Lugaz} {et~al.}(2012){Lugaz}, {Farrugia}, {Davies}, {M{\"o}stl},
  {Davis}, {Roussev}, \& {Temmer}}]{Lugaz:2012b}
{Lugaz}, N., {Farrugia}, C.~J., {Davies}, J.~A., {et~al.} 2012, Astrophys. J.,
  759, 68, \dodoi{10.1088/0004-637X/759/1/68}

\bibitem[{{Lugaz} {et~al.}(2018){Lugaz}, {Farrugia}, {Winslow}, {Al-Haddad},
  {Galvin}, {Nieves-Chinchilla}, {Lee}, \& {Janvier}}]{Lugaz:2018}
{Lugaz}, N., {Farrugia}, C.~J., {Winslow}, R.~M., {et~al.} 2018, Astrophys. J.
  Lett., 864, L7, \dodoi{10.3847/2041-8213/aad9f4}

\bibitem[{{Lugaz} {et~al.}(2010){Lugaz}, {Hernandez-Charpak}, {Roussev},
  {Davis}, {Vourlidas}, \& {Davies}}]{Lugaz:2010b}
{Lugaz}, N., {Hernandez-Charpak}, J.~N., {Roussev}, I.~I., {et~al.} 2010,
  Astrophys. J., 715, 493, \dodoi{10.1088/0004-637X/715/1/493}

\bibitem[{{Lugaz} {et~al.}(2005){Lugaz}, {Manchester}, \&
  {Gombosi}}]{Lugaz:2005b}
{Lugaz}, N., {Manchester}, W.~B., \& {Gombosi}, T.~I. 2005, Astrophys. J., 634,
  651, \dodoi{10.1086/491782}

\bibitem[{{Lugaz} {et~al.}(2017){Lugaz}, {Temmer}, {Wang}, \&
  {Farrugia}}]{Lugaz:2017}
{Lugaz}, N., {Temmer}, M., {Wang}, Y., \& {Farrugia}, C.~J. 2017, Sol. Phys.,
  292, 64, \dodoi{10.1007/s11207-017-1091-6}

\bibitem[{{Lugaz} {et~al.}(2022){Lugaz}, {Salman}, {Zhuang}, {Al-Haddad},
  {Scolini}, {Farrugia}, {Yu}, {Winslow}, {M{\"o}stl}, {Davies}, \&
  {Galvin}}]{Lugaz:2022}
{Lugaz}, N., {Salman}, T.~M., {Zhuang}, B., {et~al.} 2022, Astrophys. J., 929,
  149, \dodoi{10.3847/1538-4357/ac602f}

\bibitem[{{Lugaz} {et~al.}(2023){Lugaz}, {Lee}, {Jian}, {Allen}, {Al-Haddad},
  {Winslow}, {Lillis}, {Moestl}, {Zhuang}, {Palmerio}, {Lynch}, {Scolini},
  {Davies}, {Regnault}, {Nieves-Chinchilla}, \& {Farrugia}}]{Lugaz:2023a}
{Lugaz}, N., {Lee}, C.~O., {Jian}, L.~K., {et~al.} 2023, in Bulletin of the
  American Astronomical Society, Vol.~55, 249,
  \dodoi{10.3847/25c2cfeb.ac2c2454}

\bibitem[{{Luhmann} {et~al.}(2020){Luhmann}, {Gopalswamy}, {Jian}, \&
  {Lugaz}}]{Luhmann:2020}
{Luhmann}, J.~G., {Gopalswamy}, N., {Jian}, L.~K., \& {Lugaz}, N. 2020, Sol.
  Phys., 295, 61, \dodoi{10.1007/s11207-020-01624-0}

\bibitem[{{Luhmann} {et~al.}(2008){Luhmann}, {Curtis}, {Schroeder}, {McCauley},
  {Lin}, {Larson}, {Bale}, {Sauvaud}, {Aoustin}, {Mewaldt}, {Cummings},
  {Stone}, {Davis}, {Cook}, {Kecman}, {Wiedenbeck}, {von Rosenvinge}, {Acuna},
  {Reichenthal}, {Shuman}, {Wortman}, {Reames}, {Mueller-Mellin}, {Kunow},
  {Mason}, {Walpole}, {Korth}, {Sanderson}, {Russell}, \&
  {Gosling}}]{Luhmann:2008}
{Luhmann}, J.~G., {Curtis}, D.~W., {Schroeder}, P., {et~al.} 2008, Space Sci.
  Rev., 136, 117, \dodoi{10.1007/s11214-007-9170-x}

\bibitem[{{Martini{\'c}} {et~al.}(2022){Martini{\'c}}, {Dumbovi{\'c}},
  {Temmer}, {Veronig}, \& {Vr{\v{s}}nak}}]{Martinic:2022}
{Martini{\'c}}, K., {Dumbovi{\'c}}, M., {Temmer}, M., {Veronig}, A., \&
  {Vr{\v{s}}nak}, B. 2022, Astron. Astrophys., 661, A155,
  \dodoi{10.1051/0004-6361/202243433}

\bibitem[{{M{\"o}stl} {et~al.}(2020){M{\"o}stl}, {Weiss}, {Bailey}, {Reiss},
  {Amerstorfer}, {Hinterreiter}, {Bauer}, {McIntosh}, {Lugaz}, \&
  {Stansby}}]{Moestl:2020}
{M{\"o}stl}, C., {Weiss}, A.~J., {Bailey}, R.~L., {et~al.} 2020, Astrophys. J.,
  903, 92, \dodoi{10.3847/1538-4357/abb9a1}

\bibitem[{{M{\"u}ller} {et~al.}(2020){M{\"u}ller}, {St. Cyr}, {Zouganelis},
  {Gilbert}, {Marsden}, {Nieves-Chinchilla}, {Antonucci}, {Auch{\`e}re},
  {Berghmans}, {Horbury}, {Howard}, {Krucker}, {Maksimovic}, {Owen}, {Rochus},
  {Rodriguez-Pacheco}, {Romoli}, {Solanki}, {Bruno}, {Carlsson}, {Fludra},
  {Harra}, {Hassler}, {Livi}, {Louarn}, {Peter}, {Sch{\"u}hle}, {Teriaca}, {del
  Toro Iniesta}, {Wimmer-Schweingruber}, {Marsch}, {Velli}, {De Groof},
  {Walsh}, \& {Williams}}]{Mueller:2020}
{M{\"u}ller}, D., {St. Cyr}, O.~C., {Zouganelis}, I., {et~al.} 2020, Astron.
  Astrophys., 642, A1, \dodoi{10.1051/0004-6361/202038467}

\bibitem[{{Nykyri} {et~al.}(2023){Nykyri}, {Ma}, {Burkholder}, {Liou},
  {Cu{\'e}llar}, {Kavosi}, {Borovsky}, {Parker}, {Rosen}, {De Moudt}, {Ebert},
  {Ogasawara}, {Opher}, {Sibeck}, {Di Matteo}, {Viall}, {Wallace}, {Jorgensen},
  {Hesse}, {West}, {Adhikari}, {Argall}, {Egedal}, {Wilder}, {Broll}, {Poh},
  {Wing}, \& {Russell}}]{Nykyri:2023}
{Nykyri}, K., {Ma}, X., {Burkholder}, B., {et~al.} 2023, Frontiers in Astronomy
  and Space Sciences, 10, 1179344, \dodoi{10.3389/fspas.2023.1179344}

\bibitem[{{Owens} {et~al.}(2017){Owens}, {Lockwood}, \& {Barnard}}]{Owens:2017}
{Owens}, M.~J., {Lockwood}, M., \& {Barnard}, L.~A. 2017, Scientific Reports,
  7, 4152, \dodoi{10.1038/s41598-017-04546-3}

\bibitem[{{Owens} {et~al.}(2006){Owens}, {Merkin}, \& {Riley}}]{Owens:2006a}
{Owens}, M.~J., {Merkin}, V.~G., \& {Riley}, P. 2006, J. Geophys. Res., 111,
  3104, \dodoi{10.1029/2005JA011460}

\bibitem[{{Prise} {et~al.}(2015){Prise}, {Harra}, {Matthews}, {Arridge}, \&
  {Achilleos}}]{Prise:2015}
{Prise}, A.~J., {Harra}, L.~K., {Matthews}, S.~A., {Arridge}, C.~S., \&
  {Achilleos}, N. 2015, Journal of Geophysical Research (Space Physics), 120,
  1566, \dodoi{10.1002/2014JA020256}

\bibitem[{{Raouafi} {et~al.}(2023){Raouafi}, {Matteini}, {Squire}, {Badman},
  {Velli}, {Klein}, {Chen}, {Matthaeus}, {Szabo}, {Linton}, {Allen}, {Szalay},
  {Bruno}, {Decker}, {Akhavan-Tafti}, {Agapitov}, {Bale}, {Bandyopadhyay},
  {Battams}, {Ber{\v{c}}i{\v{c}}}, {Bourouaine}, {Bowen}, {Cattell},
  {Chandran}, {Chhiber}, {Cohen}, {D'Amicis}, {Giacalone}, {Hess}, {Howard},
  {Horbury}, {Jagarlamudi}, {Joyce}, {Kasper}, {Kinnison}, {Laker}, {Liewer},
  {Malaspina}, {Mann}, {McComas}, {Niembro-Hernandez}, {Nieves-Chinchilla},
  {Panasenco}, {Pokorn{\'y}}, {Pusack}, {Pulupa}, {Perez}, {Riley},
  {Rouillard}, {Shi}, {Stenborg}, {Tenerani}, {Verniero}, {Viall}, {Vourlidas},
  {Wood}, {Woodham}, \& {Woolley}}]{Raouafi:2023}
{Raouafi}, N.~E., {Matteini}, L., {Squire}, J., {et~al.} 2023, Space Sci. Rev.,
  219, 8, \dodoi{10.1007/s11214-023-00952-4}

\bibitem[{{Regnault} {et~al.}(2023){Regnault}, {Al-Haddad}, {Lugaz},
  {Farrugia}, {Yu}, {Zhuang}, \& {Davies}}]{Regnault:2023b}
{Regnault}, F., {Al-Haddad}, N., {Lugaz}, N., {et~al.} 2023, Astrophys. J.

\bibitem[{{Richardson} \& {Cane}(2010)}]{Richardson:2010}
{Richardson}, I.~G., \& {Cane}, H.~V. 2010, Sol. Phys., 264, 189,
  \dodoi{10.1007/s11207-010-9568-6}

\bibitem[{{Robbrecht} \& {Berghmans}(2004)}]{Robbrecht:2004}
{Robbrecht}, E., \& {Berghmans}, D. 2004, Astron. Astrophys., 425, 1097,
  \dodoi{10.1051/0004-6361:20041302}

\bibitem[{{Rodriguez} {et~al.}(2011){Rodriguez}, {Mierla}, {Zhukov}, {West}, \&
  {Kilpua}}]{Rodriguez:2011}
{Rodriguez}, L., {Mierla}, M., {Zhukov}, A.~N., {West}, M., \& {Kilpua}, E.
  2011, Sol. Phys., 270, 561, \dodoi{10.1007/s11207-011-9784-8}

\bibitem[{{Salman} {et~al.}(2021){Salman}, {Lugaz}, {Winslow}, {Farrugia},
  {Jian}, \& {Galvin}}]{Salman:2021}
{Salman}, T.~M., {Lugaz}, N., {Winslow}, R.~M., {et~al.} 2021, Astrophys. J.,
  921, 57, \dodoi{10.3847/1538-4357/ac11f3}

\bibitem[{{Scolini} {et~al.}(2023){Scolini}, {Winslow}, {Lugaz}, \&
  {Poedts}}]{Scolini:2023}
{Scolini}, C., {Winslow}, R.~M., {Lugaz}, N., \& {Poedts}, S. 2023, Astrophys.
  J., 944, 46, \dodoi{10.3847/1538-4357/aca893}

\bibitem[{{St.~Cyr} {et~al.}(2000{\natexlab{a}}){St.~Cyr}, {Mesarch},
  {Maldonado}, {Folta}, {Harper}, {Davila}, \& {Fisher}}]{StCyr:2000b}
{St.~Cyr}, O.~C., {Mesarch}, M.~A., {Maldonado}, H.~M., {et~al.}
  2000{\natexlab{a}}, J. Atmos. Solar-Terr. Phys., 62, 1251,
  \dodoi{10.1016/S1364-6826(00)00069-9}

\bibitem[{{St.~Cyr} {et~al.}(2000{\natexlab{b}}){St.~Cyr}, {Plunkett},
  {Michels}, {Paswaters}, {Koomen}, {Simnett}, {Thompson}, {Gurman}, {Schwenn},
  {Webb}, {Hildner}, \& {Lamy}}]{StCyr:2000}
{St.~Cyr}, O.~C., {Plunkett}, S.~P., {Michels}, D.~J., {et~al.}
  2000{\natexlab{b}}, J. Geophys. Res., 105, 18169,
  \dodoi{10.1029/1999JA000381}

\bibitem[{{Szabo}(2005)}]{Szabo:2005}
{Szabo}, A. 2005, Adv. Space Res., 35, 61, \dodoi{10.1016/j.asr.2004.09.013}

\bibitem[{{Thernisien}(2011)}]{Thernisien:2011}
{Thernisien}, A. 2011, 194, 33, \dodoi{10.1088/0067-0049/194/2/33}

\bibitem[{{Thernisien} {et~al.}(2009){Thernisien}, {Vourlidas}, \&
  {Howard}}]{Thernisien:2009}
{Thernisien}, A., {Vourlidas}, A., \& {Howard}, R.~A. 2009, Sol. Phys., 256,
  111, \dodoi{10.1007/s11207-009-9346-5}

\bibitem[{von Forstner(2021)}]{gcs_johan}
von Forstner, J. L.~F. 2021, johan12345/gcs\_python: Release 0.2.2, 0.2.2,
  Zenodo, \dodoi{10.5281/zenodo.5084818}

\bibitem[{{Wang} {et~al.}(2004){Wang}, {Shen}, {Wang}, \& {Ye}}]{Wang:2004}
{Wang}, Y., {Shen}, C., {Wang}, S., \& {Ye}, P. 2004, Sol. Phys., 222, 329,
  \dodoi{10.1023/B:SOLA.0000043576.21942.aa}

\bibitem[{{Winslow} {et~al.}(2021){Winslow}, {Lugaz}, {Scolini}, \&
  {Galvin}}]{Winslow:2021}
{Winslow}, R.~M., {Lugaz}, N., {Scolini}, C., \& {Galvin}, A.~B. 2021,
  Astrophys. J., 916, 94, \dodoi{10.3847/1538-4357/ac0821}

\bibitem[{{Winslow} {et~al.}(2016){Winslow}, {Lugaz}, {Schwadron}, {Farrugia},
  {Yu}, {Raines}, {Mays}, {Galvin}, \& {Zurbuchen}}]{Winslow:2016}
{Winslow}, R.~M., {Lugaz}, N., {Schwadron}, N.~A., {et~al.} 2016, J. Geophys.
  Res. Space Phys., 121, 6092, \dodoi{10.1002/2015JA022307}

\bibitem[{{Xue} {et~al.}(2005){Xue}, {Wang}, \& {Dou}}]{Xue:2005}
{Xue}, X.~H., {Wang}, C.~B., \& {Dou}, X.~K. 2005, J. Geophys. Res., 110, 8103,
  \dodoi{10.1029/2004JA010698}

\bibitem[{{Yashiro} {et~al.}(2004){Yashiro}, {Gopalswamy}, {Michalek},
  {St.~Cyr}, {Plunkett}, {Rich}, \& {Howard}}]{Yashiro:2004}
{Yashiro}, S., {Gopalswamy}, N., {Michalek}, G., {et~al.} 2004, J. Geophys.
  Res., 109, A07105, \dodoi{10.1029/2003JA010282}

\bibitem[{{Yu} {et~al.}(2016){Yu}, {Farrugia}, {Galvin}, {Lugaz}, {Luhmann},
  {Simunac}, \& {Kilpua}}]{Yu:2016}
{Yu}, W., {Farrugia}, C.~J., {Galvin}, A.~B., {et~al.} 2016, J. Geophys. Res.,
  121, 5005, \dodoi{10.1002/2016JA022642}

\bibitem[{{Zhao} {et~al.}(2017){Zhao}, {Feng}, {Feng}, \& {Li}}]{Zhao:2017}
{Zhao}, X.~H., {Feng}, X.~S., {Feng}, H.~Q., \& {Li}, Z. 2017, Astrophys. J.,
  849, 79, \dodoi{10.3847/1538-4357/aa8e49}

\bibitem[{{Zurbuchen} \& {Richardson}(2006)}]{Zurbuchen:2006}
{Zurbuchen}, T.~H., \& {Richardson}, I.~G. 2006, Space Sci. Rev., 123, 31,
  \dodoi{10.1007/s11214-006-9010-4}

\end{thebibliography}
\end{document}